\documentclass[twocolumn,twocolappendix]{aastex7}

\usepackage{soul}
\newcommand{\yw}{\color{black}}
\newcommand{\yc}{\color{black}}

\begin{document}

\title{The Hourglass-shaped Magnetic Fields and Dust Filaments in the HH 211 Protostellar Envelope}

\author[0009-0004-9279-780X]{Youngwoo Choi}
\affiliation{Department of Physics and Astronomy, Seoul National University, Gwanak-ro 1, Gwanak-gu, Seoul, 08826, Korea}
\email{}

\author[0000-0003-4022-4132]{Woojin Kwon}
\affiliation{Department of Earth Science Education, Seoul National University, Gwanak-ro 1, Gwanak-gu, Seoul, 08826, Korea}

\affiliation{SNU Astronomy Research Center, Seoul National University, Gwanak-ro 1, Gwanak-gu, Seoul, 08826, Korea}

\affiliation{The Center for Educational Research, Seoul National University, 1 Gwanak-ro, Gwanak-gu, Seoul 08826, Republic of Korea}
\email{}

\author[0000-0002-4540-6587]{Leslie W. Looney}
\affiliation{Department of Astronomy, University of Illinois, 1002 West Green Street, Urbana, IL 61801, USA}
\email{}

\author[0000-0003-3017-4418]{Ian W. Stephens}
\affiliation{Department of Earth, Environment, and Physics, Worcester State University, Worcester, MA 01602, USA}
\email{}

\author[0000-0002-7402-6487]{Zhi-Yun Li}
\affiliation{Astronomy Department, University of Virginia, 530 McCormick Rd., Charlottesville, Virginia 22903, USA}
\email{}

\author[0000-0002-8942-1594]{Floris F.S. van der Tak}
\affiliation{SRON Netherlands Institute for Space Research, Landleven 12, 9747 AD Groningen, The Netherlands}

\affiliation{Kapteyn Astronomical Institute, University of Groningen, P.O. Box 800, 9700 AV Groningen, The Netherlands}
\email{}

\author[0000-0002-6195-0152]{John J. Tobin}
\affiliation{National Radio Astronomy Observatory, 520 Edgemont Road, Charlottesville, VA, 22093, USA}
\email{}

\correspondingauthor{Woojin Kwon}
\email{wkwon@snu.ac.kr}

\begin{abstract}

Magnetic fields influence the structure and evolution of protostellar systems, thus understanding their role is essential for probing the earliest stages of star formation. We present ALMA Band 3 and 6 polarized continuum {\yc observations} at $\sim$0.5$^{\prime \prime}$ resolution toward the Class 0 protostellar system HH 211. Three dust filaments ($\sim$4000 au in length) are found in the HH 211 protostellar envelope, two of which are aligned with core-scale ($\sim$10,000 au) magnetic fields detected by previous JCMT observations. This result suggests that the formation of the dust filaments may be influenced by magnetic fields. {\yc In the inner envelope ($\sim$1000 au), we detect a clear hourglass-shaped magnetic field morphology near the protostar and toroidal fields along the outflow directions.} {\yc We also estimate the line-of-sight–averaged temperature and column density distributions in the inner envelope and find that the temperature is higher in the east, while the column density is enhanced in the southern and western regions.} {\yw The southern dense regions of the inner envelope may trace either outflow cavity walls, due to their alignment with the outflow, or possible infalling channels in the midplane, given the close correspondence between the observed magnetic fields and the predicted infall trajectories.}

\end{abstract}


\section{Introduction} \label{sec:intro}

Stars form through the gravitational collapse of molecular cloud cores \citep{Shu_1987,Larson_2003,McKee_2007}. Traditionally, it has been considered that core materials collapse isotropically into the central part of the core, forming a protostar \citep[e.g.,][]{Shu_1977}. However, recent high-resolution observations have revealed numerous asymmetric accretion channels around various young protostellar systems, known as streamers \citep{Pineda_2023}. Streamers have been observed around young stellar objects from the youngest Class 0 to older Class II disks, indicating gravitational collapse occurs in asymmetrical ways during the star formation process \citep[e.g.,][]{Le_Gouellec_2019,Yen_2019,Pineda_2020,Cabedo_2021,Thieme_2022,Murillo_2022,Garufi_2022,Valdivia_Mena_2022,Valdivia_Mena_2023,Leej_2023,Kido_2023,Aso_2023,Flores_2023,Cacciapuoti_2024,Tychoniec_2024,Codella_2024,Riaz_2024,Hales_2024}.

Several numerical simulations have shown that streamers are naturally caused by turbulence in the parent cloud cores \citep{Walch_2010,Seifried_2013,Seifried_2015}. In the presence of turbulence, the gravitational collapse of cores occurs asymmetrically, as observed in the simulations \citep[e.g.,][]{Matsumoto_2017,Lam_2019,Hennebelle_2020,Tu_2023,Heigl_2024}. Additionally, since star-forming cores are embedded within the filamentary structures of molecular clouds \citep{Andre_2014}, streamers could also be generated through interactions with the surrounding medium \citep[e.g.,][]{Kuffmeier_2017,Bate_2018,Kuffmeier_2023}. Collisions between cores or filaments present another possible mechanism for forming asymmetric structures around protostellar systems \citep{Yano_2024}.

Another important feature of star-forming cores is that they are threaded with ambient magnetic fields \citep{Crutcher_2012}. As core material collapses toward the center, the magnetic fields are dragged inward, forming a classical hourglass shape \citep{Mestel_1966,Mouschovias_1976,Galli_1993,Basu_1994,Myers_2018,Myers_2020}. Hourglass magnetic fields are observed around several high-mass \citep[e.g.,][]{Schleuning_1998,Girart_2009,Tang_2009,Qiu_2014,Li_2015} and low-mass protostellar systems \citep[e.g.,][]{Girart_2006,Stephens_2013,Maury_2018,Kwon_2019}. In addition, the magnetic fields near protostars are twisted by rotational motions, forming toroidal fields along the jet directions \citep{Pudritz_2019}. These magnetic fields are expected to play a significant role in the formation of protostellar disks, jets, and outflows \citep[e.g.,][]{Blandford_1982,Machida_2006,Li_2013}.

HH 211 is a Class 0 protostellar system located in the IC 348 star-forming region of the Perseus molecular cloud complex. The distance to HH 211 is about 300 pc \citep{Ortiz_Le_n_2018, Zucker_2018}, and its bolometric temperature is 27 K \citep{Tobin_2016}. HH 211 was first discovered through the observation of its molecular hydrogen jets \citep{McCaughrean_1994}, and subsequent observations revealed a highly collimated jet and outflow with distinct knot structures \citep{Lee_2009,Jhan_2016,Jhan_2021,Ray_2023}. Possible toroidal magnetic fields along the jet have also been detected through SiO polarization observations \citep{Lee_2018a}. A thick, nearly edge-on Keplerian disk, approximately 20 au in size, is embedded within the protostellar envelope \citep{Lee_2018b, Lee_2023}. The total mass of the central protostar and embedded disk is estimated to be $\sim$0.08 $M_{\odot}$ \citep{Lee_2019}. Infalling and rotating motions are observed in the HH 211 protostellar envelope \citep{Tanner_2011,Lee_2019}, and the mass of the HH 211 core ($\sim$10,000 au scale) is about 2.5 $M_{\odot}$ \citep{Yen_2023, Choi_2024}.

In this paper, we study the magnetic fields and physical properties within the HH 211 protostellar envelope. We present ALMA Band 3 ($\sim$97.5 GHz) and Band 6 ($\sim$233 GHz) polarization observations toward the HH 211 protostellar system. Observations and data reductions are described in Section \ref{sec:observations}. In Section \ref{sec:results}, we present observational results. We discuss physical properties of HH 211 in Section \ref{sec:discussions} and summarize our conclusions in Section \ref{sec:conclusion}.

\section{Observations} \label{sec:observations}

\subsection{ALMA Band 3} \label{subsec:band3}
HH 211 was observed by ALMA in Band 3 from December 14 to 26, 2017 (2017.1.01310.S, PI: Woojin Kwon). The minimum and maximum baseline lengths are about 4 $k\lambda$ and 1100 $k\lambda$, respectively. The maximum recoverable scale ($\theta_{\mathrm{MRS}}$) is about 31$^{\prime \prime}$. We used CASA to calibrate the raw data. J0522+3627 and J0510+1800 were used for bandpass and polarization calibrations during individual executions, while J0237+2848 and J0341+3352 were used to calibrate flux and phase, respectively. We also performed phase-only self-calibration using the Stokes $I$ map with solution intervals of 300 s to reduce the noise in the final Stokes $I$, $Q$, and $U$ maps. The resulting peak SNR for Stokes $I$ is 753. We made images with the CASA $tclean$ task using a Briggs weighting with robust parameters of 0.5 and 2.0. With a robust parameter of 0.5, the resulting beam size is $0.514^{\prime \prime} \times 0.337^{\prime \prime}$ (PA =$-28.9^{\circ}$), and the rms noises of final Stokes $I$, $Q$, and $U$ maps are $9.0 \times 10^{-3}$ mJy beam$^{-1}$, $5.8 \times 10^{-3}$ mJy beam$^{-1}$, and $5.8 \times 10^{-3}$ mJy beam$^{-1}$, respectively. With a robust parameter of 2.0, the resulting beam size is $0.754^{\prime \prime} \times 0.536^{\prime \prime}$ (PA =$-31.1^{\circ}$), and the rms noises of final Stokes $I$, $Q$, and $U$ maps are $1.0 \times 10^{-2}$ mJy beam$^{-1}$, $6.0 \times 10^{-3}$ mJy beam$^{-1}$, and $6.0 \times 10^{-3}$ mJy beam$^{-1}$, respectively. Polarized intensity is debiased with $PI = \sqrt{Q^2 + U^2 - \sigma_{QU}^2}$, where $\sigma_{QU}$ is the rms noise of Stokes $Q$ and $U$. The rms noises of the polarized intensity are $8.2 \times 10^{-3}$ mJy beam$^{-1}$ and $7.5 \times 10^{-3}$ mJy beam$^{-1}$ for robust parameters of 0.5 and 2.0, respectively. The Stokes $V$ parameter is not detected, with an rms noise of $5.8 \times 10^{-3}$ mJy beam$^{-1}$ for a robust parameter of 0.5.

\subsection{ALMA Band 6} \label{subsec:band6}
HH 211 was observed by ALMA in Band 6 from November 12 to 14, 2016 (2016.1.00604.S, PI: Woojin Kwon). The minimum and maximum baseline lengths are about 9 $k\lambda$ and 825 $k\lambda$, respectively. The maximum recoverable scale ($\theta_{\mathrm{MRS}}$) is about 14$^{\prime \prime}$. We used CASA to calibrate the raw data. J0238+1636 was used for flux and polarization calibration, while J0237+2848 and J0336+3218 were used to calibrate bandpass and phase, respectively. We also performed phase-only self-calibration using the Stokes $I$ map with solution intervals of 300 s, 100 s, and 30 s to reduce the noise in the final Stokes $I$, $Q$, and $U$ maps. The resulting peak SNR for Stokes $I$ is 1166. We made images with the CASA $tclean$ task using a Briggs weighting with robust parameters of 0.5 and 2.0. With a robust parameter of 0.5, the resulting beam size is $0.587^{\prime \prime} \times 0.355^{\prime \prime}$ (PA = $15.1^{\circ}$), and the rms noises of Stokes $I$, $Q$, and $U$ maps are $4.5 \times 10^{-2}$ mJy beam$^{-1}$, $1.3 \times 10^{-2}$ mJy beam$^{-1}$, and $1.3 \times 10^{-2}$ mJy beam$^{-1}$, respectively. With a robust parameter of 2.0, the resulting beam size is $0.742^{\prime \prime} \times 0.471^{\prime \prime}$ (PA = $9.9^{\circ}$), and the rms noises of final Stokes $I$, $Q$, and $U$ maps are $3.5 \times 10^{-2}$ mJy beam$^{-1}$, $1.3 \times 10^{-2}$ mJy beam$^{-1}$, and $1.3 \times 10^{-2}$ mJy beam$^{-1}$, respectively. Polarized intensity is debiased with $PI = \sqrt{Q^2 + U^2 - \sigma_{QU}^2}$, where $\sigma_{QU}$ is the rms noises of Stokes $Q$ and $U$. The rms noises of the polarized intensity are $1.9 \times 10^{-2}$ mJy beam$^{-1}$ and $1.2 \times 10^{-2}$ mJy beam$^{-1}$ for robust parameter of 0.5 and 2.0, respectively. The Stokes $V$ parameter is not detected, with an rms noise of $1.2 \times 10^{-2}$ mJy beam$^{-1}$ for a robust parameter of 0.5.

\subsection{JCMT 850 $\mu$m} \label{subsec:jcmt}

The HH 211 core was observed using the JCMT SCUBA-2 camera with the POL-2 polarimeter between July (Project ID: M17AP073, PI: Woojin Kwon) and September (Project ID: M17BP058, PI: Woojin Kwon) 2017, and between October 2019 and February 2020 as part of the BISTRO survey (Project ID: M17BL011) \citep{Ward_Thompson_2017}. We used 850 $\mu$m polarization continuum data to infer core-scale ($\sim$10,000 au) magnetic fields around HH 211. Detailed data reduction process is described in \cite{Choi_2024}. The beam size of the JCMT observations is 14.6$^{\prime \prime}$, and the rms noises of final Stokes $I$, $Q$, and $U$ maps at 850 $\mu$m are 3.3 mJy beam$^{-1}$, 2.5 mJy beam$^{-1}$, and 2.5 mJy beam$^{-1}$, respectively.

\section{Results} \label{sec:results}

\subsection{Dust Continuum and Core-scale Magnetic Field} \label{subsec:continuum intensity maps}

\begin{figure*}[ht!]
\epsscale{0.8}
\plotone{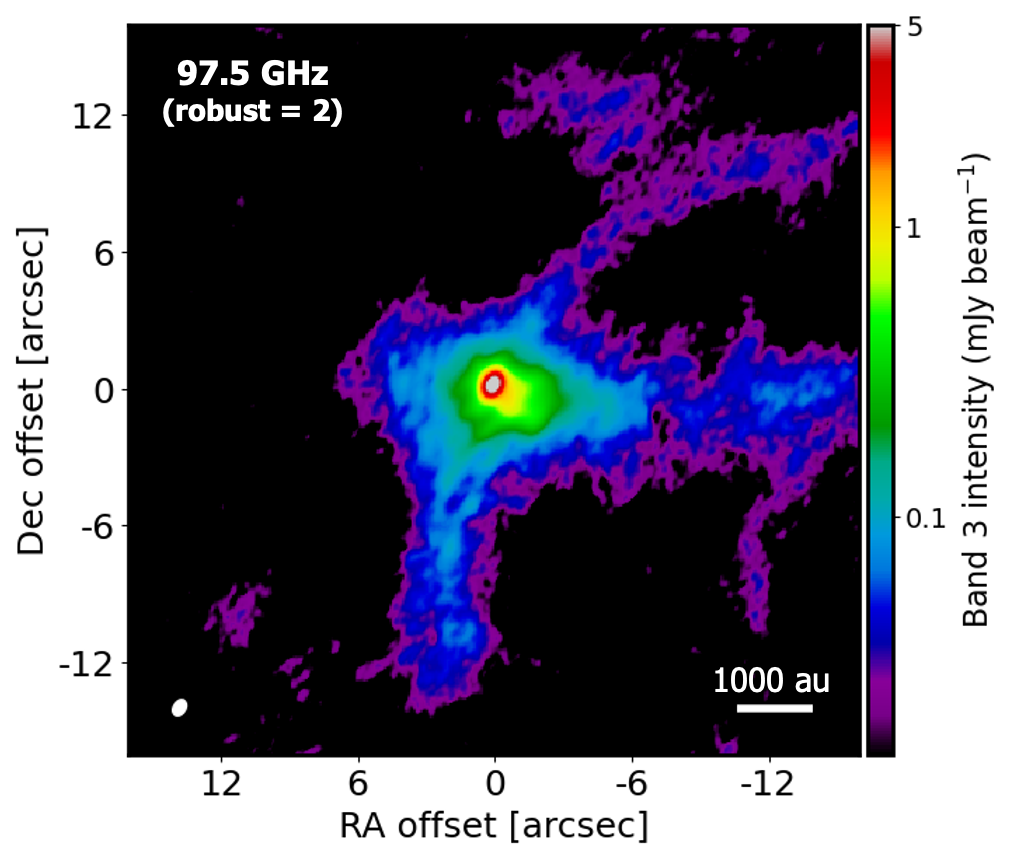}
\caption{Dust continuum around the HH 211 in ALMA Band 3 (97.5 GHz) with a robust parameter of 2.0. The synthesized beam and spatial scale bar are shown in the bottom left and right corners, respectively. {\yc The central coordinates are RA = 03$^{\mathrm{h}}$43$^{\mathrm{m}}$56.800$^{\mathrm{s}}$, Dec = +32$^\circ$00$'$50.00$''$.}
\label{fig:cont_band3}}
\end{figure*}

\begin{figure*}[ht!]
\epsscale{1.15}
\plotone{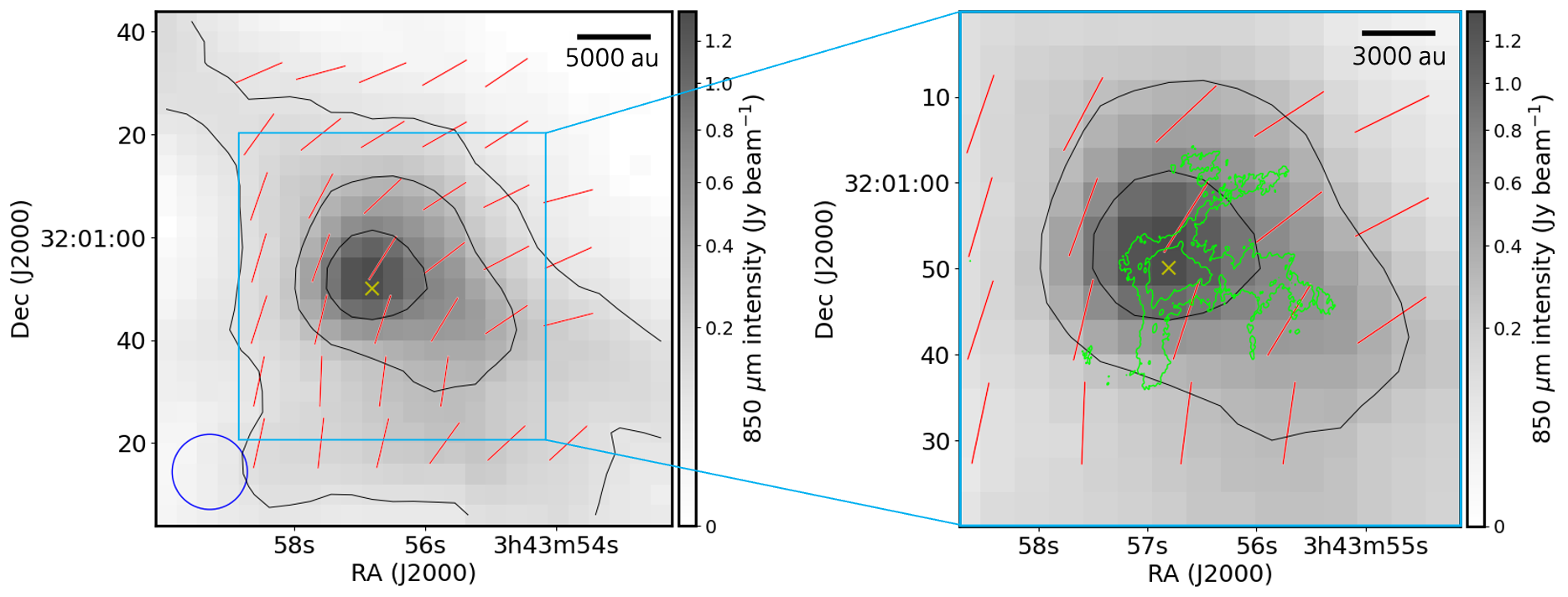}
\caption{Left: The core-scale ($\sim$10,000 au) magnetic field orientations around the HH 211 observed with the JCMT. The greyscale shows the 850 $\mu$m intensity, with contour levels at 0.09, 0.3, and 0.8 Jy beam$^{-1}$. The beam size of the JCMT observation is shown in the bottom left of the panel. The cross marker indicates the position of the central protostar \citep{Lee_2018b,Lee_2023}. Right: A zoomed-in view of the left panel overlaid with ALMA Band 3 observations. The green lines represent 0.025 mJy beam$^{-1}$ and 0.08 mJy beam$^{-1}$ levels in the ALMA Band 3 observations with a robust parameter of 2.0. Spatial scale bars are shown in the top right corner of both panels.
\label{fig:jcmt}}
\end{figure*}

\begin{figure*}[ht!]
\epsscale{1.1}
\plotone{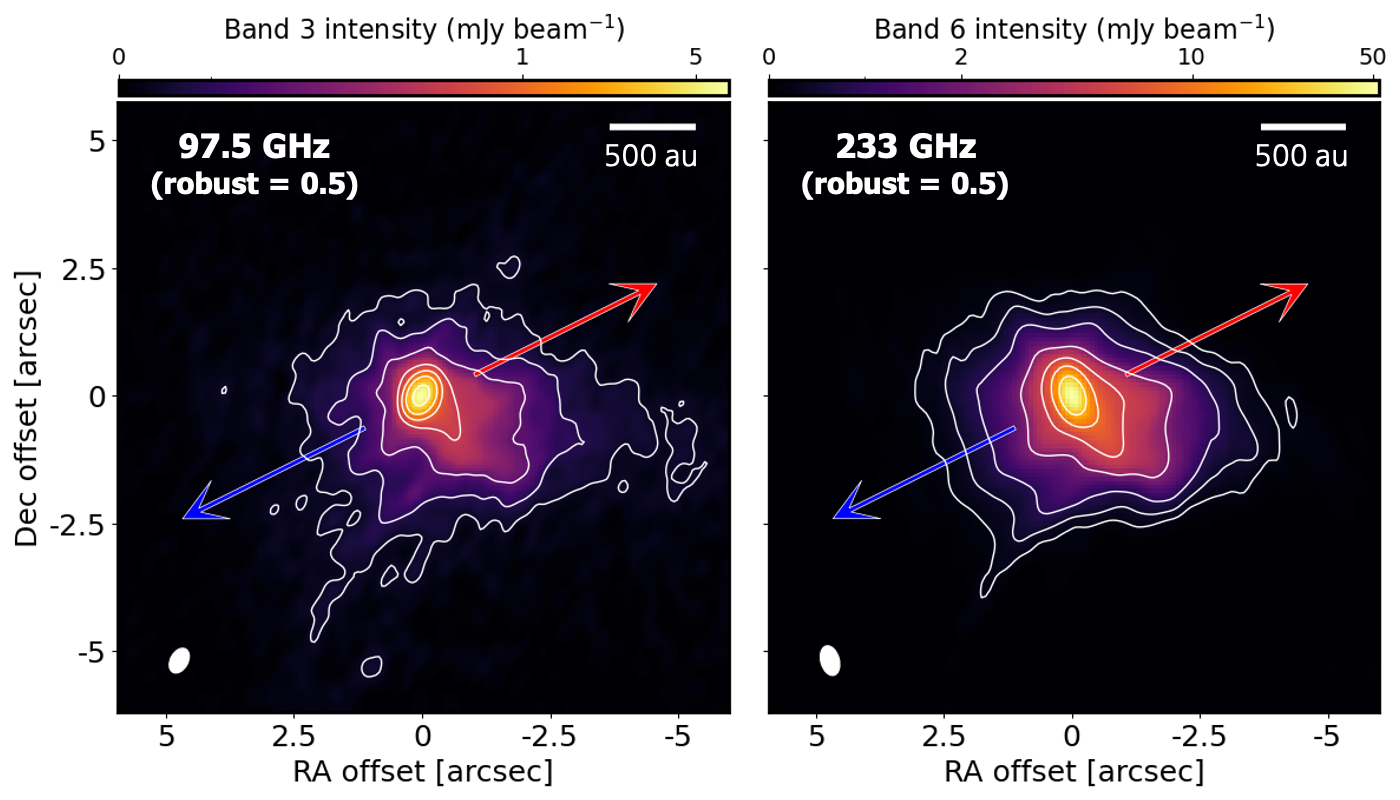}
\caption{Dust continuum around the HH 211 in Band 3 (Left) and Band 6 (Right) with a robust parameter of 0.5. The contour levels are 5, 10, 20, 50, 100, 200, and 500 times the rms noise levels, where the rms noises are 0.009 mJy beam$^{-1}$ in Band 3 and 0.045 mJy beam$^{-1}$ in Band 6. The synthesized beam sizes are shown in the bottom left of both panels. The red and blue arrows indicate the red-shifted and blue-shifted jet directions, respectively \citep{McCaughrean_1994}. Spatial scale bars are shown in the top right corner of both panels.
\label{fig:continuum}}
\end{figure*}

Figure \ref{fig:cont_band3} shows dust continuum toward the HH 211 envelope in Band 3 (97.5 GHz) with a robust parameter of 2.0. We detect three dust filaments, each about 4000 au in scale, in the HH 211 envelope. These filaments are located in the southern, western, and northwestern regions of the HH 211 envelope, all directed toward the central embedded protostar. The intensity of the filaments is the strongest in the southern filament and the faintest in the northwestern filament. For comparison, dust continuum emission in Band 6 with a robust parameter of 2.0 is presented in Appendix \ref{sec:app}.

The left panel of Figure \ref{fig:jcmt} shows the core-scale ($\sim$10,000 au) magnetic field orientations around HH 211 observed by JCMT with the POL-2 polarimeter. The right panel of Figure \ref{fig:jcmt} shows a zoomed-in view of the core-scale magnetic fields overlaid with ALMA Band 3 continuum. Notably, the southern and northwestern filaments are well aligned with the core-scale magnetic fields. This alignment suggests that the formation of southern and northwestern filaments is likely influenced by the magnetic fields. If these filaments are infalling mass streams, they could be guided by the magnetic fields. In contrast, the western filament is almost perpendicular to the core-scale magnetic fields. Instead, it is directed toward the asymmetrically extended part of the core, as shown in the right panel of Figure \ref{fig:jcmt}.

Figure \ref{fig:continuum} shows dust continuum toward the HH 211 envelope in Band 3 (97.5 GHz) and Band 6 (233 GHz) with a robust parameter of 0.5. Since the size of the embedded disk is about 20 au \citep{Lee_2018b,Lee_2023}, which is smaller than our beam size, the disk is not resolved in our observations. At both frequencies, there is a narrow filamentary structure in the southern part of the envelope, which is connected to the larger southern dust filament shown in Figure \ref{fig:cont_band3}.

In addition, the dust continuum is asymmetrically extended toward the southwest direction near the central protostar. \cite{Lee_2009} first reported this feature with SMA observations, suggesting the possibility of a binary companion at $0.3^{\prime \prime}$ to the southwest of HH 211. However, high-resolution ALMA observations did not show any secondary source at the predicted position \citep{Lee_2018b, Lee_2019}. This asymmetrically extended envelope might be due to the asymmetric density distributions, as will be discussed in a Section \ref{sec:discussions}.

Another feature of the HH 211 envelope is that it is flattened along the east-west direction on a 2000 au scale. The position angle of the minor axis of the flattened envelope is about 165$^{\circ}$. The position angle of the outflow direction is about 116$^{\circ}$, which is nearly perpendicular to the position angle of the embedded disk \citep{Lee_2018b,Segura-Cox_2018}, as indicated in Figure \ref{fig:continuum}. The outflow direction is tilted by 50$^{\circ}$ relative to the flattened envelope direction on the 2000 au scale. Since the outflow direction, a proxy of the rotation axis, is not aligned with the the short axis of the flattened structure, this may not be the result of rotational motions. Instead, the core-scale ($\sim$10,000 au) magnetic field at the position of the protostar has a position angle of about 155$^{\circ}$, which is more aligned with the flattening axis within 10$^{\circ}$. This suggests that the flattened envelope may be a pseudodisk, possibly formed by the preferential mass accumulation along the ambient magnetic field lines.

\subsection{Magnetic Field Morphology} \label{subsec:magnetic field morphology}

\begin{figure*}[ht!]
\epsscale{1.15}
\plotone{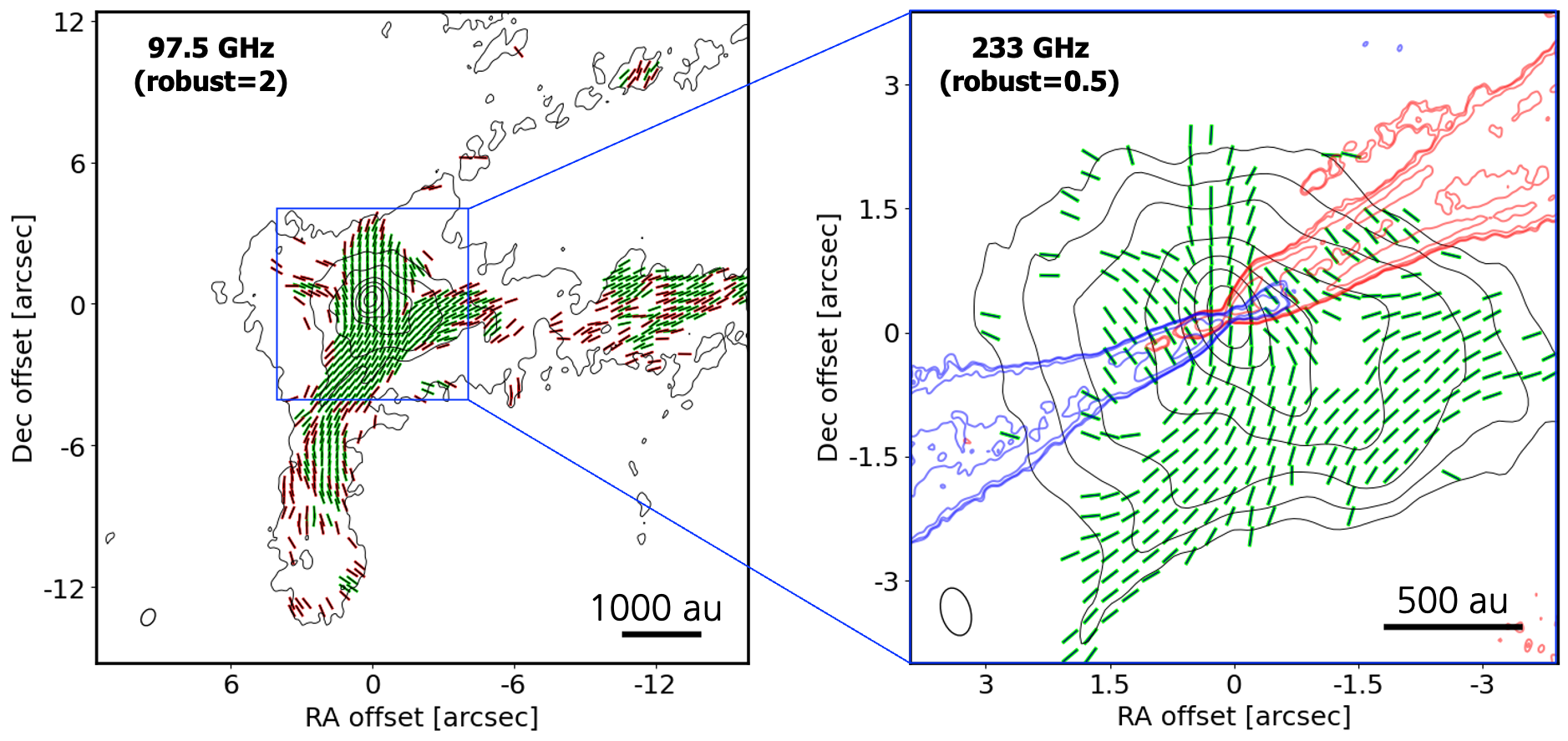}
\caption{Left: Inferred magnetic field orientations around the HH 211 from Band 3 polarized continuum with a robust parameter of 2.0. The line segments represent plane-of-sky magnetic field directions, with green for $PI > 2\sigma_{PI}$, and red for $1.5\sigma_{PI}< PI < 2\sigma_{PI}$. The black lines are the Band 3 dust continuum with a robust parameter of 2, showing 3, 10, 20, 50, 100, 200, and 500 times the rms noise level. Right: Inferred magnetic field orientations around HH 211 from Band 6 polarized continuum with a robust parameter of 0.5. Green segments represent plane-of-sky magnetic field directions with $PI > 3\sigma_{PI}$. The blue and red contours indicate CO (2-1) integrated intensities in velocity ranges from 0 to 8 and from 10 to 18 km s$^{-1}$, respectively \citep{Jhan_2021}. The blue and red contour levels are 0.03, 0.035, 0.05, 0.1, and 0.2 Jy beam$^{-1}$ km s$^{-1}$. The black lines are the Band 6 dust continuum with a robust parameter of 0.5, showing 5, 10, 20, 50, 100, 200, and 500 times the rms noise level. The synthesized beam size and spatial scale bar are shown in the bottom left and right of both panels, respectively. 
\label{fig:B_field}}
\end{figure*}

Figure \ref{fig:B_field} shows the inferred plane-of-sky magnetic field orientations toward the HH 211 envelope from Band 3 and Band 6 polarization observations. The plane-of-sky magnetic field orientations are inferred from $90^{\circ}$ rotation of polarization directions. We selected the Band 3 image with a robust parameter of 2.0 and the Band 6 image with a robust parameter of 0.5, as larger-scale magnetic fields are more effectively detected in Band 3, whereas smaller-scale fields are more clearly revealed in Band 6. For Band 3 observations, we represent half-vectors satisfying $I > 3\sigma_{I}$ and $PI > 1.5\sigma_{PI}$, while for Band 6, the criteria are $I > 3\sigma_{I}$ and $PI > 3\sigma_{PI}$. Magnetic field vectors are sampled at intervals of 0.24$^{\prime\prime}$ and 0.32$^{\prime\prime}$ for robust parameters of 0.5 and 2.0, respectively, which are approximately half the beam size of the corresponding observations. Red and blue contours in the right panels indicate red-shifted and blue-shifted CO (2-1) integrated intensities, respectively \citep{Jhan_2021}. For comparison, magnetic field orientations from the Band 3 observations with a robust parameter of 0.5 and Band 6 with a robust parameter of 2.0 are presented in Appendix \ref{sec:app}.

The left panel of Figure \ref{fig:B_field} shows the magnetic field orientations within the southern and western filaments. In the southern filament, the magnetic fields are aligned with the filament direction, consistent with the core-scale magnetic fields shown in Figure \ref{fig:jcmt}. In the case of the western filament, the filament is oriented in the east-west direction. The magnetic fields within the western filament are almost aligned with the filament but slightly tilted toward the northwest. Since the core-scale magnetic fields at the western filament are oriented toward the northwest, as shown in Figure \ref{fig:jcmt}, the magnetic field directions within the western filament are likely inherited from the core-scale magnetic fields. If the southern and western filaments are infalling streams toward the central protostars, the observed magnetic field directions might result from the fields dragged by these streams, aligning the field direction with the infalling direction. The magnetic fields within the northwestern filament are not clearly detected in our Band 3 observations and also not in Band 6.

The right panel of Figure \ref{fig:B_field} shows the magnetic field orientations in the HH 211 inner envelope ($\sim500$ au). The magnetic fields in the southern and western parts of the envelope are connected to the magnetic fields in the southern and western filaments shown in the left panel of Figure \ref{fig:B_field}. Near the central protostar, we detect clear U-shaped pinched magnetic fields around the blue-shifted outflow axis, as well as a hint of pinched fields around the red-shifted outflow. The pinched fields associated with the red-shifted outflow are more clearly seen in the Band 6 image with a robust parameter of 2.0, shown in the right panel of Figure \ref{fig:B_field_appendix}. This represents so-called hourglass-shaped magnetic fields, indicating that infalling materials are well coupled to the magnetic fields, dragging the fields toward the central region. At the center of the HH 211 envelope, the inferred magnetic fields from both Band 3 and Band 6 observations are oriented in the north-south direction, consistent with previous Band 7 (358 GHz) observations \citep{Lee_2019,Yen_2023}.

The magnetic field is perpendicular to the outflow directions in the outflow regions. This field morphology is observed in both the red-shifted and blue-shifted outflows, suggesting the presence of toroidal magnetic fields. Previous SiO polarization observations also detected possible toroidal fields on the blue-shifted side of the jet \citep{Lee_2018a}. SiO observations trace the inside of the outflow, where dust grains are destroyed by shocks, while dust emission primarily originates from the envelope material in front of and behind the outflow. Therefore, outflow rotation could generate toroidal fields within the outflow regions and also induce the rotation of the envelope, leading to the formation of toroidal fields in the envelope.

\subsection{Polarized Intensity and Polarization Fraction} \label{subsec:polarization fraction}

\begin{figure*}[ht!]
\epsscale{1.15}
\plotone{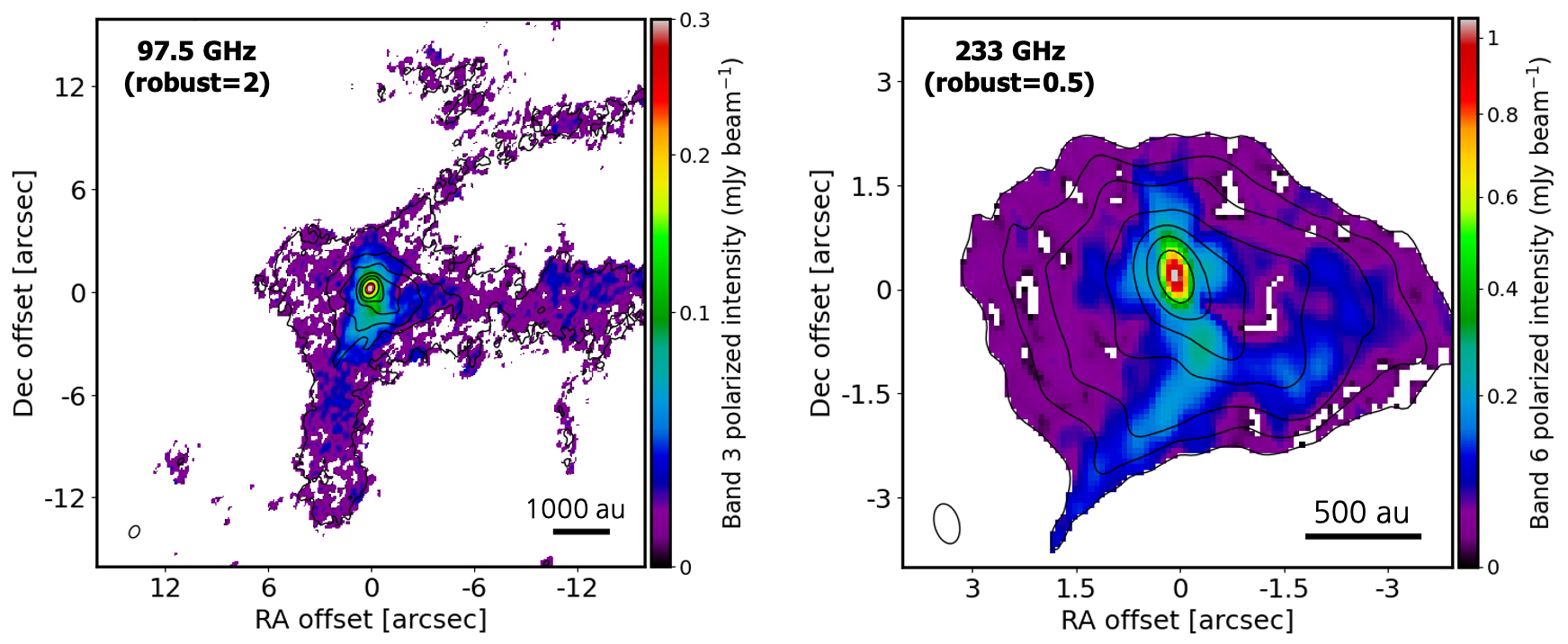}
\caption{Polarized intensity around the HH 211 envelope in Band 3 with a robust parameter of 2.0 (Left) and in Band 6 with a robust parameter of 0.5 (Right). The black lines represent Stokes $I$ intensity levels with 3 (Left) / 5 (Right), 10, 20, 50, 100, 200, and 500 times the rms noise level in each observation. The synthesized beam size and spatial scale bar are shown in the bottom left and right of both panels, respectively.
\label{fig:pol_I}}
\end{figure*}

\begin{figure*}[ht!]
\epsscale{1.15}
\plotone{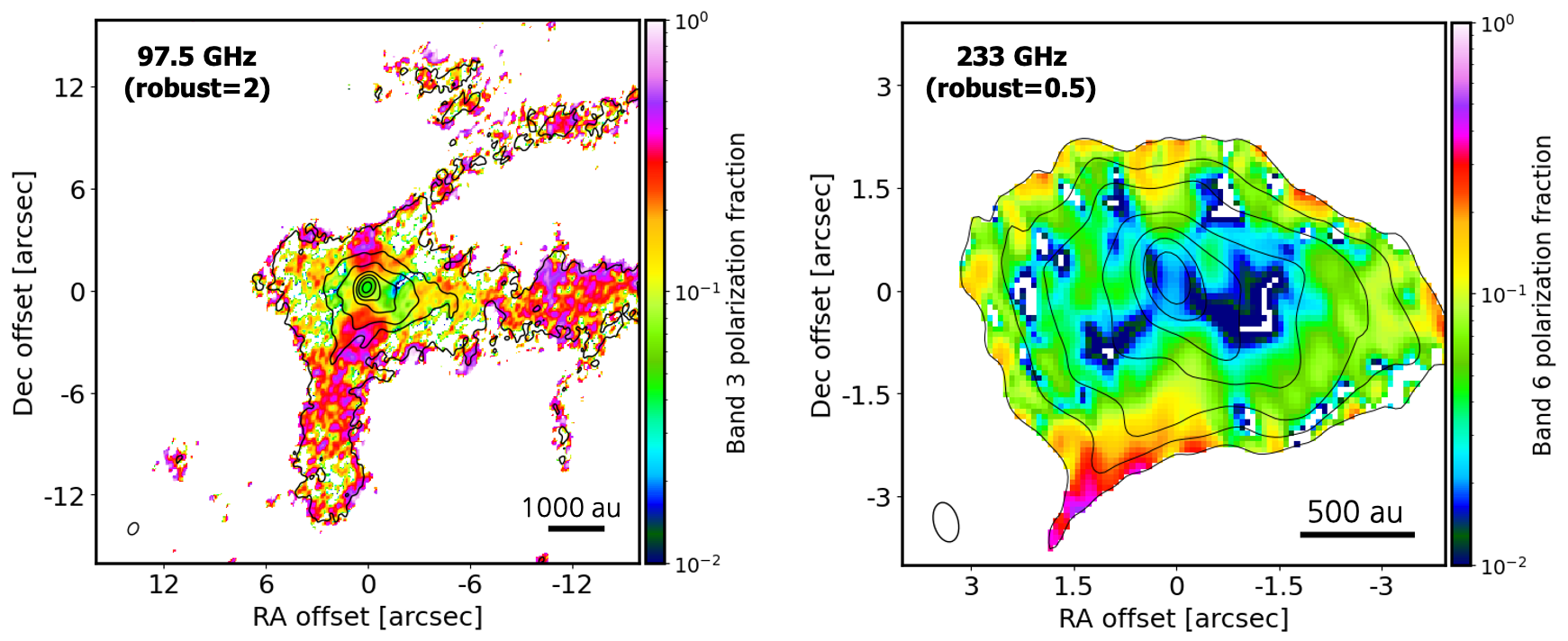}
\caption{Polarization fraction around the HH 211 envelope in Band 3 with a robust parameter of 2.0 (Left) and in Band 6 with a robust parameter of 0.5 (Right). The black lines represent Stokes $I$ intensity levels with 3 (Left) / 5 (Right), 10, 20, 50, 100, 200, and 500 times the rms noise level in each observation. The synthesized beam size and spatial scale bar are shown in the bottom left and right of both panels, respectively.
\label{fig:pol_frac}}
\end{figure*}

Figure \ref{fig:pol_I} and \ref{fig:pol_frac} show the polarized intensity ($PI$) and polarization fraction toward the HH 211 envelope in Band 3 with a robust parameter of 2.0 (left panel) and in Band 6 with a robust parameter of 0.5 (right panel). The polarized intensity and polarization fraction from the Band 3 observations with a robust parameter of 0.5 and Band 6 with a robust parameter of 2.0 are presented in Appendix \ref{sec:app}. In Band 6, the polarized intensity is high in the southern and western parts of the envelope, where it connects with the larger filaments. 

The polarization fraction is calculated with $PI/I$, where $I$ represents the Stokes $I$ parameter and $PI$ represent the polarized intensity. In Band 3, the polarization fraction is high in the northern and southern regions near the central protostar. An X-shaped depolarization pattern appears in the Band 6 observations (see the blue regions in the right panel of Figure \ref{fig:pol_frac}). These depolarization regions might represent the boundaries between toroidal fields and pinched hourglass fields, where the plane-of-sky magnetic field directions change abruptly. 

\subsection{Spectral Index} \label{subsec:spectral index}

\begin{figure*}[t!]
\epsscale{1.15}
\plotone{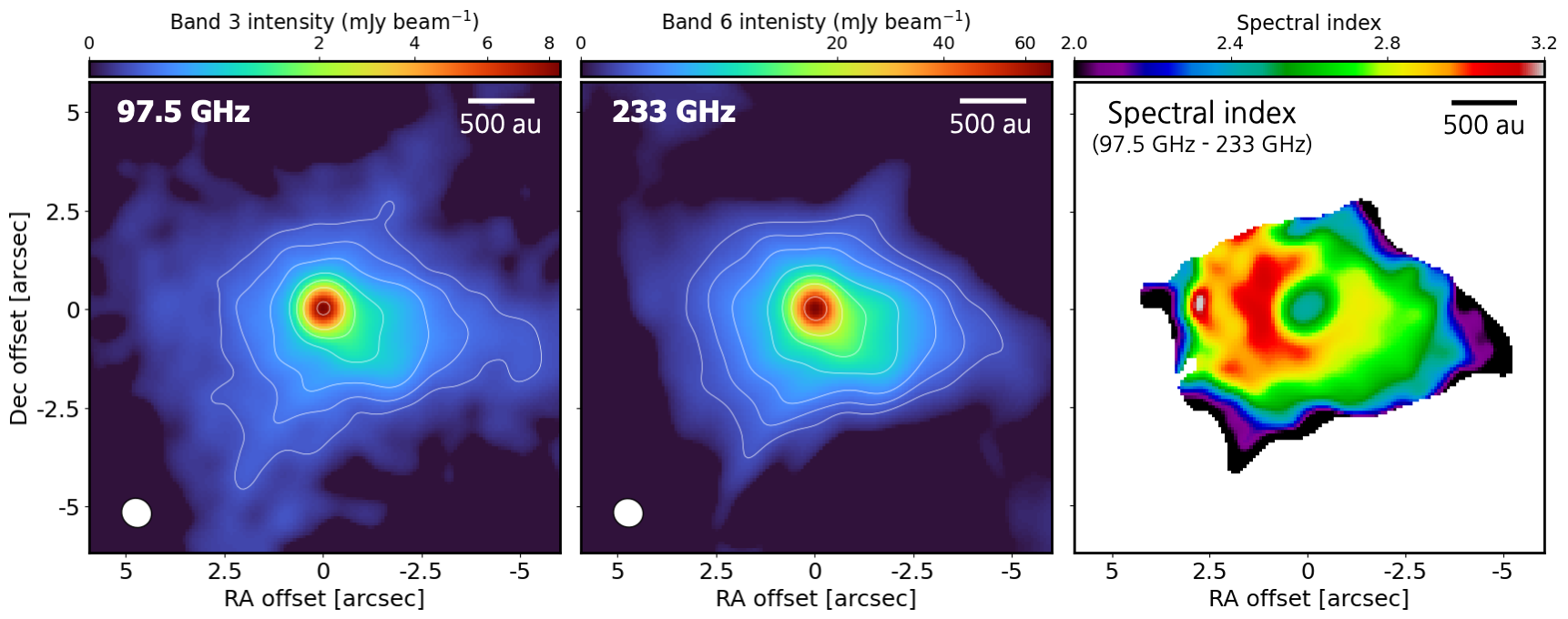}
\caption{Dust continuum images with similar beam sizes in Band 3 (Left) and Band 6 (Middle). The beam sizes are matched by adjusting the uv-coverage of the Band 3 and Band 6 observations. The resulting beam sizes are 0.774$^{\prime \prime}$ $\times$ 0.727$^{\prime \prime}$ (PA = $35.9^{\circ}$) in Band 3, and 0.759$^{\prime \prime}$ $\times$ 0.717$^{\prime \prime}$ (PA = $23.9^{\circ}$) in Band 6. Contour levels are at 5, 10, 20, 50, 100, 200, and 500 times the rms noise level in each observation. The spectral index map is made with these images (Right).
\label{fig:spectral_index}}
\end{figure*}

We investigate the spectral index between Band 3 and 6 observations to study the physical properties of the HH 211 envelope, which will be discussed in Section \ref{sec:discussions}. The spectral index of dust continuum emission between two frequencies can be calculated as follows:
\begin{equation}
\alpha = \frac{\mathrm{log} \hspace{0.08cm} (I_{\nu_{1}}/I_{\nu_{2}})}{\mathrm{log} \hspace{0.08cm} (\nu_{1}/\nu_{2})},
\label{eq:alpha}
\end{equation}
where $\nu_{1}$ and $\nu_{2}$ are 233 and 97.5 GHz, respectively. Before estimating the spectral index using Equation \ref{eq:alpha}, we matched the $uv$-coverage of the Band 3 and Band 6 observations by adjusting \texttt{uvtaper} and \texttt{uvrange} parameters during the CASA \texttt{tclean} task. In this process, the minimum $uv$-distance is set to 9 $k\lambda$. Figure \ref{fig:spectral_index} presents the Band 3 and Band 6 dust continuum images after matching their $uv$-coverage, and the corresponding spectral index map. The spectral index map is shown where both Band 3 and Band 6 intensities are larger than 3 times their rms noise levels. The spectral index has a decreasing hole structure in the central region where the spectral index becomes 2.5. The spectral index gradually increases outwardly from the center, then decreases again outwardly. Also, the spectral index shows an asymmetric structure with high values in the east and smaller values in the west.

\section{Analysis and Discussion} \label{sec:discussions}

\subsection{Temperature and Density Distributions} \label{subsec:temp_dens}

To better understand the observed spectral index structure around HH 211, we formalize the expression for the spectral index. The dust continuum without background radiation is calculated as follows:
\begin{equation}
I_{\nu} = \int_{0}^{\tau_{\nu}}{B_{\nu}(T)e^{-\tau^{\prime}_{\nu}} d\tau^{\prime}_{\nu}},
\label{eq:radiation}
\end{equation}
where $I_{\nu}$ is the intensity, $\tau_{\nu}$ is the optical depth, and $B_{\nu}(T)$ is the blackbody radiation at temperature $T$. The optical depth $\tau_{\nu}$ is calculated as follows:
\begin{eqnarray}
\tau_{\nu} &=& \kappa_{\nu}^{\prime} \Sigma_{\mathrm{d}}, \\
&=& \kappa_{\nu}^{\prime} \int{\rho_{\mathrm{d}}} \hspace{0.1cm} ds, \\
&=& \mu m_{\mathrm{H}} \kappa_{\nu} \int{n(\mathrm{H}_2)} \hspace{0.1cm} ds, \\
&=& \mu m_{\mathrm{H}} \kappa_{\nu} N (\mathrm{H}_2), 
\label{eq:opt_dep}
\end{eqnarray}
where $\kappa_{\nu}^{\prime}$ is the dust opacity, $\kappa_{\nu}$ is the dust opacity divided by gas-to-dust ratio, $\Sigma_{\mathrm{d}}$ is the dust mass column density, $\rho_{\mathrm{d}}$ is the dust density, $\mu$ is the mean molecular weight, $m_\mathrm{H}$ is the mass of a hydrogen atom, $n(\mathrm{H}_2)$ is the number density of molecular hydrogen and $N(\mathrm{H}_2)$ is the column density of molecular hydrogen. Dust opacity has the form of $\kappa_{\nu} \propto \nu^{\beta}$ in the far-IR and mm wavelengths \citep{Hildebrand_1983,Draine_2006}, where $\beta$ is the dust opacity spectral index.

Equation \ref{eq:radiation} is simplified through 
\begin{eqnarray}
I_{\nu} &=& \int_{0}^{\tau_{\nu}}{B_{\nu}(T)e^{-\tau^{\prime}_{\nu}} d\tau^{\prime}_{\nu}}, \\
&=& B_{\nu}(\overline{T}_{\mathrm{los}}) \int_{0}^{\tau_{\nu}}{e^{-\tau^{\prime}_{\nu}} d\tau^{\prime}_{\nu}}, \\
&=& B_{\nu}(\overline{T}_{\mathrm{los}})(1-e^{-\tau_{\nu}}),
\label{eq:rad_modified}
\end{eqnarray}
where $\overline{T}_{\mathrm{los}}$ is the line-of-sight averaged temperature, assuming uniform temperature along the line-of-sight. The definition of $\overline{T}_{\mathrm{los}}$ becomes as follows:
\begin{equation}
B_{\nu}(\overline{T}_{\mathrm{los}}) = \frac{\int_{0}^{\tau_{\nu}}{B_{\nu}(T)e^{-\tau^{\prime}_{\nu}} d\tau^{\prime}_{\nu}}}{\int_{0}^{\tau_{\nu}}{e^{-\tau^{\prime}_{\nu}} d\tau^{\prime}_{\nu}}}.
\label{eq:los_temp}
\end{equation}

\begin{figure}[t!]
\epsscale{1.15}
\plotone{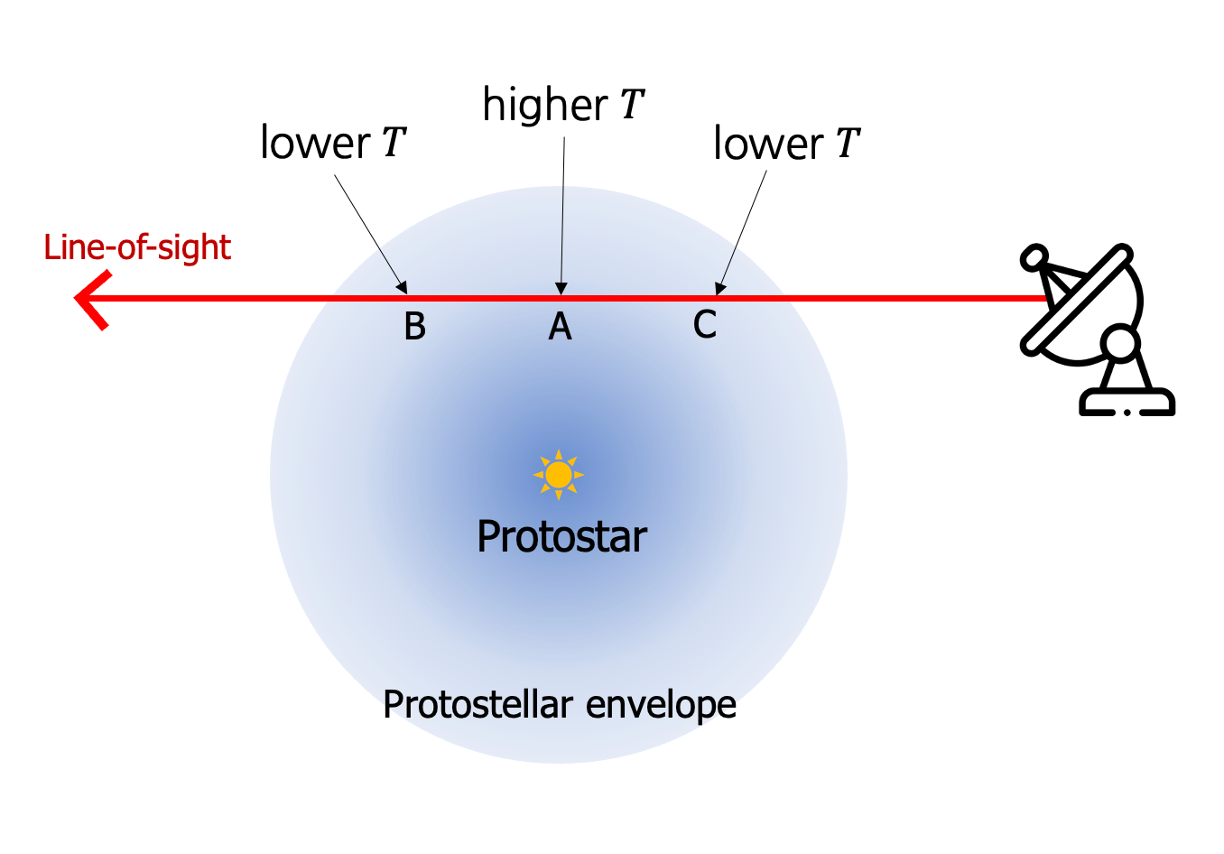}
\caption{A cartoon illustrating the observation of a protostellar envelope. The dust continuum traces not only the material closest to the protostar ($A$), but also the cooler background ($B$) and foreground ($C$) material, resulting in a line-of-sight averaged temperature lower than the temperature at location $A$.
\label{fig:los_temp}}
\end{figure}

Figure \ref{fig:los_temp} illustrates a cartoon depicting the observation of a protostellar envelope. Generally, the temperature decreases outward from the protostar, leading to a non-uniform temperature along the line-of-sight. The region closest to the protostar ($A$) is warmer than the background ($B$) and foreground ($C$). Since dust emission from the protostellar envelope is optically thin at millimeter wavelengths, the dust continuum traces both the warm material at position $A$ and the cooler material in the background and foreground. As a result, the line-of-sight averaged temperature is lower than the temperature at location $A$.

\begin{figure*}[ht!]
\epsscale{1.15}
\plotone{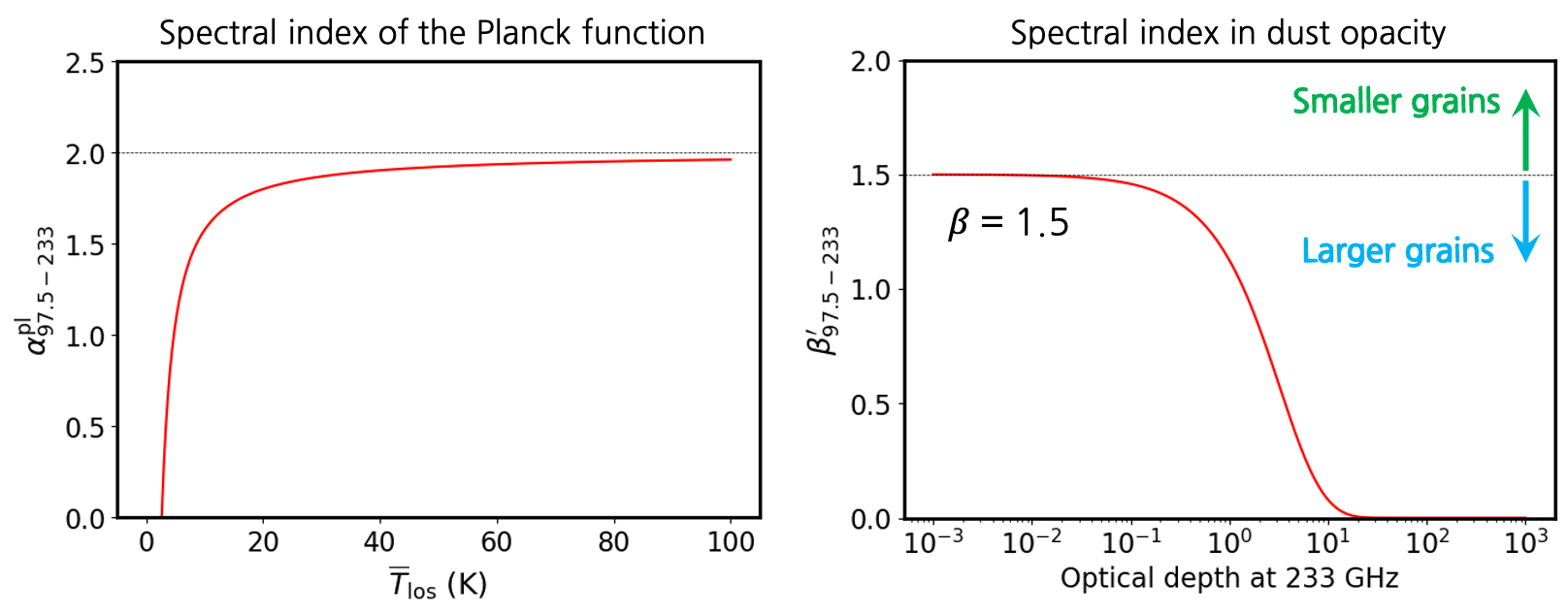}
\caption{Spectral index components between 97.5 GHz and 233 GHz frequencies. Left: Spectral index of the Planck function as a function of temperature. It converges to 2 at high temperatures. Right: $\beta^{\prime}(\tau_{\nu})$ as a function of optical depth at 233 GHz when $\beta$ is 1.5. It converges to $\beta$ in the optically thin limit and 0 in the optically thick limit.
\label{fig:alpha}}
\end{figure*}

By substituting Equation \ref{eq:rad_modified} into Equation \ref{eq:alpha}, the spectral index of the dust continuum emission can be calculated as the sum of two components \citep[e.g.,][]{Silsbee_2022}:
\begin{eqnarray}
\alpha &=& \alpha_{\mathrm{pl}}(\overline{T}_{\mathrm{los}}) + \frac{\beta \tau_{\nu_1}}{e^{\tau_{\nu_1}}-1}, \\
&=& \alpha_{\mathrm{pl}}(\overline{T}_{\mathrm{los}}) + \beta^{\prime}(\tau_{\nu_1}),
\label{eq:alpha2}
\end{eqnarray}
where $\alpha_{\mathrm{pl}}(\overline{T}_{\mathrm{los}})$ is the spectral index of the Planck function at temperature $T = \overline{T}_{\mathrm{los}}$. Figure \ref{fig:alpha} shows the two components of the spectral index between 97.5 GHz and 233 GHz from Equation \ref{eq:alpha2}. The left panel of Figure \ref{fig:alpha} represents $\alpha_{\mathrm{pl}}(\overline{T}_{\mathrm{los}})$ as a function of the line-of-sight averaged temperature ($\overline{T}_{\mathrm{los}}$), while the right panel indicates $\beta^{\prime}(\tau_{\nu_1})$ as a function of the optical depth ($\tau_{\nu_1}$), assuming a dust opacity spectral index ($\beta$) of 1.5 at the protostellar envelope. Note that the dust opacity spectral index in protostellar envelopes is generally observed to be lower than in the diffuse ISM \citep{Kwon_2009, Galametz_2019, Cacciapuoti_2025}. Without contamination from other emissions, such as free-free emission or anomalous microwave emission (AME), the spectral index is determined by three factors: the line-of-sight averaged temperature ($\overline{T}_{\mathrm{los}}$), the optical depth ($\tau_{\nu_1}$), and the dust opacity spectral index ($\beta$). As the line-of-sight averaged temperature increases, the spectral index of the Planck function $\alpha_{\mathrm{pl}}(\overline{T}_{\mathrm{los}})$ also increases, asymptotically approaching 2 at high temperatures. This value corresponds to the Rayleigh–Jeans limit of the Planck function at the observing frequencies. In the optically thin regime, the spectral index associated with dust opacity $\beta^{\prime}(\tau_{\nu_1})$ approaches $\beta$ and decreases as the optical depth increases. In the optically thick limit, the spectral index for dust opacity becomes 0. The dust opacity spectral index ($\beta$) can vary based on dust grain size, with smaller grains resulting in higher $\beta$ values, and larger grains resulting in lower $\beta$ \citep{Draine_2006}. In the Rayleigh-Jeans and the optically thin limits, Equation \ref{eq:alpha2} reduces to $\alpha = 2 + \beta$.

The outwardly decreasing spectral index beyond a radius of about 2$^{\prime \prime}$ seen in Figure \ref{fig:spectral_index} is likely caused by the decreasing temperature, as the optical depth decreases outward. The dust opacity spectral index, $\beta$, is also not expected to decrease outward. This suggests that the temperature of the HH 211 envelope is low enough that the Rayleigh-Jeans approximation does not hold. The inward decrease of the spectral index in the central region is likely due to the emission from the optically thick embedded disk. This disk emission might be smoothed out by the beam pattern, increasing optical depth in the central region.

Since the optical depth depends on the dust opacity and the column density (refer to Equation \ref{eq:opt_dep}), the dust continuum emission is governed by three parameters: the line-of-sight averaged temperature, the dust opacity, and the column density (refer to Equation \ref{eq:rad_modified}). Therefore, by assuming a dust opacity value, the dust continuum depends on the line-of-sight averaged temperature and the column density. To estimate these parameters for the HH 211 envelope, we take $\mu = 2.86$, $\beta=1.5$, $\kappa_{\nu}=0.01$ at 233 GHz assuming the gas-to-dust ratio of 100 \citep{Ossenkopf_1994}. We assume that $\beta$ is constant across the region. Then, we solve Equation \ref{eq:rad_modified} using the Band 3 and 6 images in Figure \ref{fig:spectral_index}, and calculate the line-of-sight averaged temperature and the column density simultaneously.

\begin{figure*}[ht!]
\epsscale{1.15}
\plotone{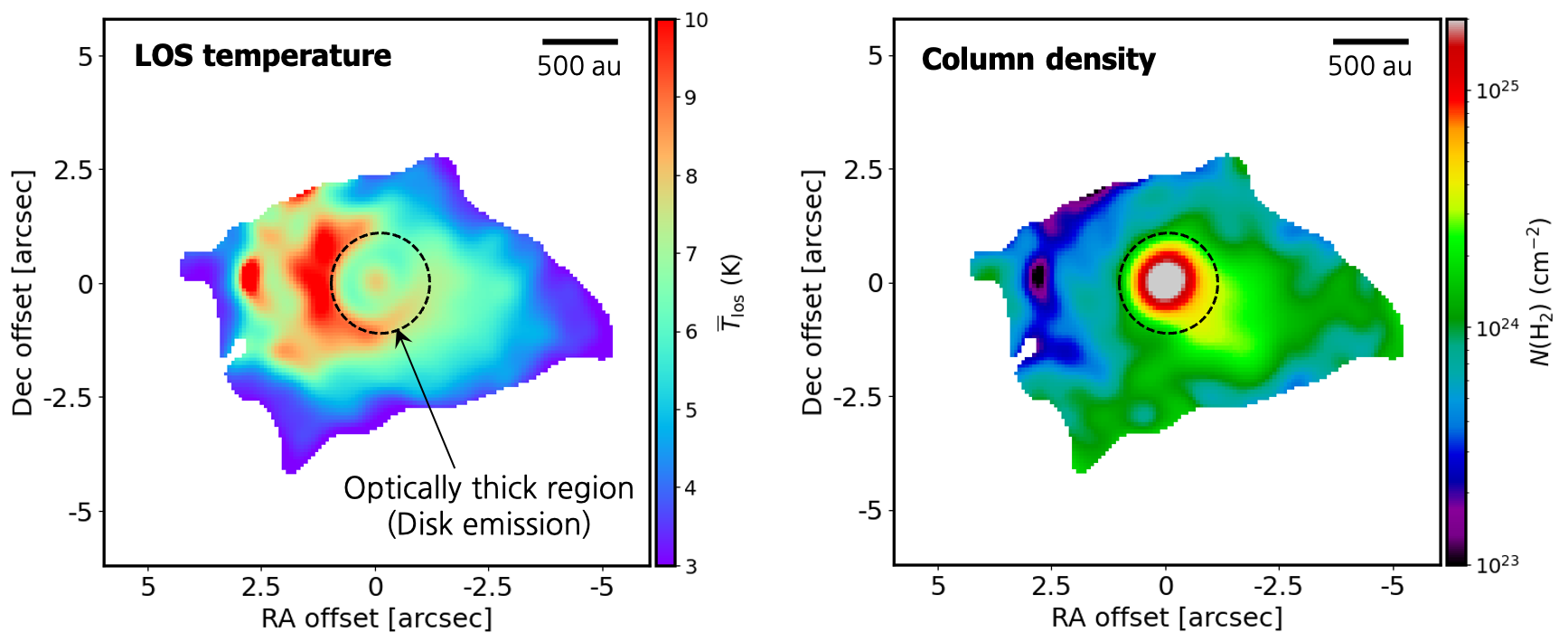}
\caption{The estimated line-of-sight averaged temperature (Left) and column density (Right) of the HH 211 envelope. Note that calculations are uncertain in the central region due to the embedded disk emission. 
\label{fig:temp_dens}}
\end{figure*}

Figure \ref{fig:temp_dens} shows the estimated line-of-sight averaged temperature and column density of the HH 211 envelope. Note that the line-of-sight averaged temperature is generally lower than the temperature at the projected radial distance from the protostar (refer to Figure \ref{fig:los_temp}). Estimates in the central region are uncertain due to contamination from the embedded disk emission, which increases the optical depth and affects the assumed $\beta$ value. Consequently, the estimated line-of-sight temperature in the center is also uncertain. The line-of-sight temperature pattern is similar to the spectral index, as the spectral index is sensitive to the temperature. The estimated line-of-sight temperature is high in the east and low in the west, with a decrease outward, reaching $\sim$4 K at the boundary. This unexpectedly low temperature may be due to the fact that the continuum emission is resolved out on large scales, which could behave differently at the two bands, even though we match the minimum $uv$-distance. Another possibility is that it could result from the assumed $\beta$ value. If $\beta$ is as low as 1.2, the estimated line-of-sight temperature at the boundary would increase to about 6 K.

The estimated column density also shows an asymmetric structure. The column density is enhanced in the southwestern part of the envelope, corresponding to the asymmetrically extended emission observed in the dust continuum. The estimated column density also reveals narrow channels in the south and the west, consistent with the dust filament locations observed in the dust continuum. This suggests that the dust filaments observed in the dust continuum are likely regions of high density rather than high temperature. In Appendix \ref{sec:modeling}, we demonstrate that our analysis can reliably distinguish between temperature and density structures from the dust continuum. If the dust filaments are infalling toward the protostar, our results indicate that accretion in HH 211 occurs preferentially through the southern and western channels, though this should be confirmed with high-resolution spectral line observations. We emphasize that the inferred line-of-sight temperature and column density patterns remain consistent regardless of the assumptions of dust opacity and $\beta$ values. Although different dust opacity values will affect the resulting line-of-sight temperature and column density values, their overall patterns do not change.

\subsection{Outflow Cavity Walls or Infalling Streams?} \label{subsec:kinematics}

\begin{figure}[t!]
\epsscale{1.15}
\plotone{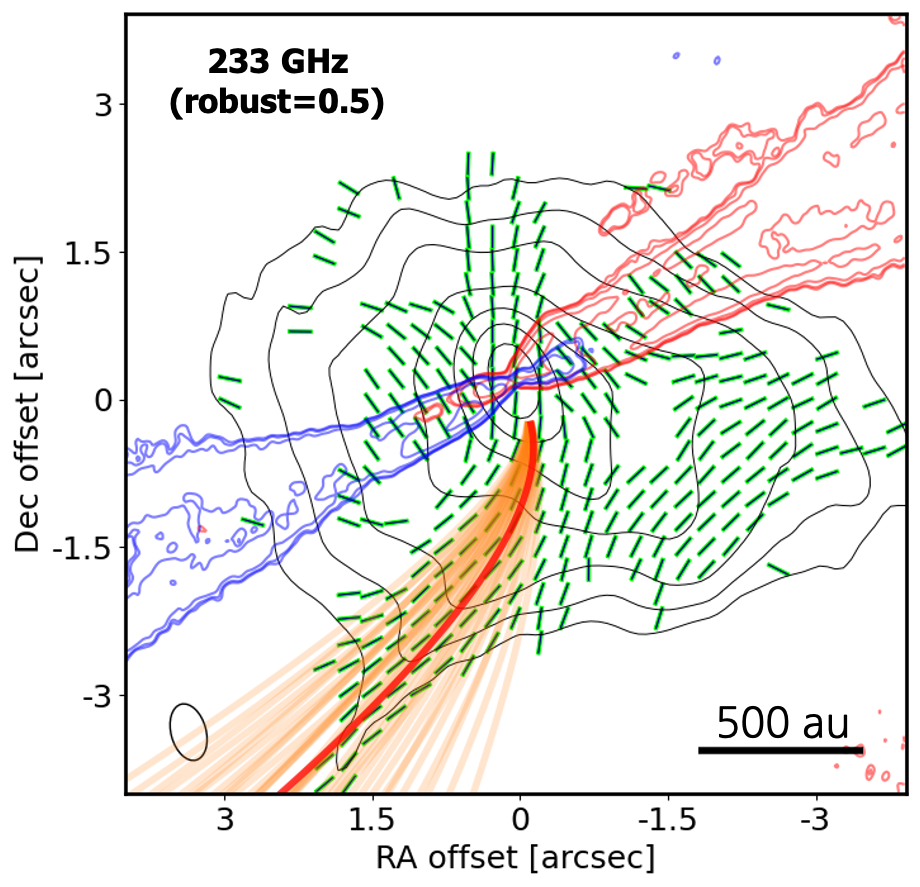}
\caption{{\yw CMU model trajectories in the southern region of the HH 211 envelope. The red line shows the case with $\theta_{0}=90^{\circ}, \phi_{0}=90^{\circ}$, while the orange lines represent models with $80^{\circ}<\theta_{0}<90^{\circ}$ and $90^{\circ}<\phi_{0}<95^{\circ}$. All other features are the same as in the right panel of Figure \ref{fig:B_field}.}
\label{fig:infall_traj}}
\end{figure}

{\yw On a $\sim$1000 au scale, the HH 211 envelope shows an enhanced column density toward the south and west (Figure \ref{fig:temp_dens}). The dust continuum emission in both Band 3 and Band 6 reveals an elongated filamentary structure in the southern region (Figure \ref{fig:continuum}), while the magnetic field orientations (Figure \ref{fig:B_field}) are well aligned with this filament. One interpretation is that the southern filament traces the outflow cavity walls. Similar cavity wall filaments have also been reported in other Class 0 protostellar envelopes \citep[e.g.,][]{Maury_2018,Kwon_2019,Le_Gouellec_2019}. Alternatively, the southern filament might not be physically related to the outflow. Its apparent alignment with the outflow could simply result from a projection effect. Instead, the southern feature may correspond to a dense infalling stream that supplies material to the central protostar.}

{\yc To investigate possible infall motions {\yw along the southern filament}, we adopt the CMU (Cassen–Moosman–Ulrich) model \citep{Ulrich_1976, Cassen_1981}, which describes ballistic, parabolic trajectories resulting from the gravitational collapse of a rotating system. The model assumes an initial solid-body rotation and a central point mass, and it neglects pressure forces and magnetic fields. Infalling streamer trajectories in protostellar envelopes have been successfully identified using the CMU model in several previous studies \citep[e.g.,][]{Thieme_2022, Kido_2023, Aso_2023}. The model contains six free parameters: the central protostellar mass ($M_*$), the centrifugal radius ($R_c$), the disk inclination ($i$), the position angle of the disk (PA), and the initial polar and azimuthal angles ($\theta_0$, $\phi_0$) of the infalling particles. These parameters determine the shape of the parabolic trajectory and the velocity field along the streamline. For details on calculating the CMU model trajectory, refer to Appendix \ref{sec:cmu}.}

{\yc To infer parabolic infall trajectories around HH 211, we adopt a central mass of $M_* + M_{\text{disk}} = 0.08$ $M_\odot$ \citep{Lee_2018b, Lee_2019}, a centrifugal radius of $R_c = 42$ au \citep{Lee_2019}, and disk inclination and position angle of $i = 98^\circ$ and PA = $64^\circ$, respectively \citep{Lee_2023}. For the definitions of inclination and position angle used in this study, refer to Appendix \ref{sec:cmu}. {\yw With these four parameters fixed, we search for CMU model trajectories that reproduce the southern filamentary structures traced by the dust continuum emission and aligned magnetic field lines.} {\yw Figure \ref{fig:infall_traj} shows the inferred CMU model trajectories along the southern filament. The red curve corresponds to the trajectory with initial angles $\theta_{0}=90^\circ$ and $\phi_{0}=90^\circ$, while the orange curves represent models with $80^\circ < \theta_{0} < 90^\circ$ and $90^\circ < \phi_{0} < 95^\circ$. The inferred trajectories align well with the observed magnetic field lines. This can be interpreted as magnetic field lines being dragged inward by infalling motions, as shown in previous simulations \citep[e.g.,][]{Mignon_2021,Tu_2023}. Since the inferred trajectories are located near the equatorial plane ($\theta_0=90^{\circ}$), the filament's apparent association with the outflow could be a projection effect. Upon careful inspection, we find no other sets of initial angles ($\theta_0$, $\phi_0$) that yield trajectories consistent with both the observed dust filamentary structures and the curvature of the magnetic field, apart from those originating near the midplane region ($80^\circ < \theta_0 < 95^\circ$). We note that one cannot arbitrarily obtain infalling trajectories that reproduce any curved field lines simply by varying the initial angles ($\theta_0$, $\phi_0$), given the fixed physical conditions ($M_*$, $R_c$) and viewing geometry ($i$, PA) of the system. The trajectories shown in Figure \ref{fig:infall_traj} are derived under the well-constrained physical parameters of HH 211, and thus support the interpretation that the magnetic field lines could be dragged inward by infall motions.}}

\subsection{Polarized Filaments in Protostellar Envelopes} \label{subsec:magnetized_streamers}

The polarized intensity shown in the right panel of Figure \ref{fig:pol_I} {\yw also} highlights filamentary structures in the south. Similar polarized filamentary structures have been observed in various Class 0 protostellar envelopes, such as Ser-emb-8 \citep{Hull_2017}, IRAS 16293 \citep{Sadavoy_2018}, B335 \citep{Maury_2018}, L1448 IRS2 \citep{Kwon_2019}, and Serpens SMM1 \citep{Le_Gouellec_2019}. \cite{Le_Gouellec_2023} argued that these polarized filaments are associated with outflow cavity walls, suggesting they are produced by enhanced radiation from the protostar onto the outflow cavity walls, which aligns dust grains within the wall. {\yw In HH 211, the polarized filaments also coincide with the outflow cavity walls in the plane of the sky.} {\yw However, our results further indicate that these filaments are associated with the high-density regions, rather than high-temperature regions.} Notably, the magnetic field directions observed within the polarized filaments are almost parallel to the filament orientations and directed toward the central protostars \citep[e.g.,][]{Hull_2017,Sadavoy_2018,Maury_2018,Kwon_2019,Le_Gouellec_2019, Le_Gouellec_2023}. {\yw For the southern polarized filament in HH 211, this field morphology is consistent with the infall trajectories in the midplane shown in Figure \ref{fig:infall_traj}, suggesting that the magnetic fields on $\sim$1000 au scale may be dragged by infalling motions.} {\yw \citet{Le_Gouellec_2023,Le_Gouellec_2023_2} reported that polarized filaments are also detected in the equatorial planes of protostellar systems, where ordered magnetic fields can produce highly linear polarization.} {\yw Taken together, the southern polarized filament in HH 211 could therefore correspond either to outflow cavity walls or to dense infalling streams within the midplane.}

\subsection{Mass Infalling Rate along the Southern Filament} \label{subsec:infall_rate}

\begin{figure}[t!]
\epsscale{1.15}
\plotone{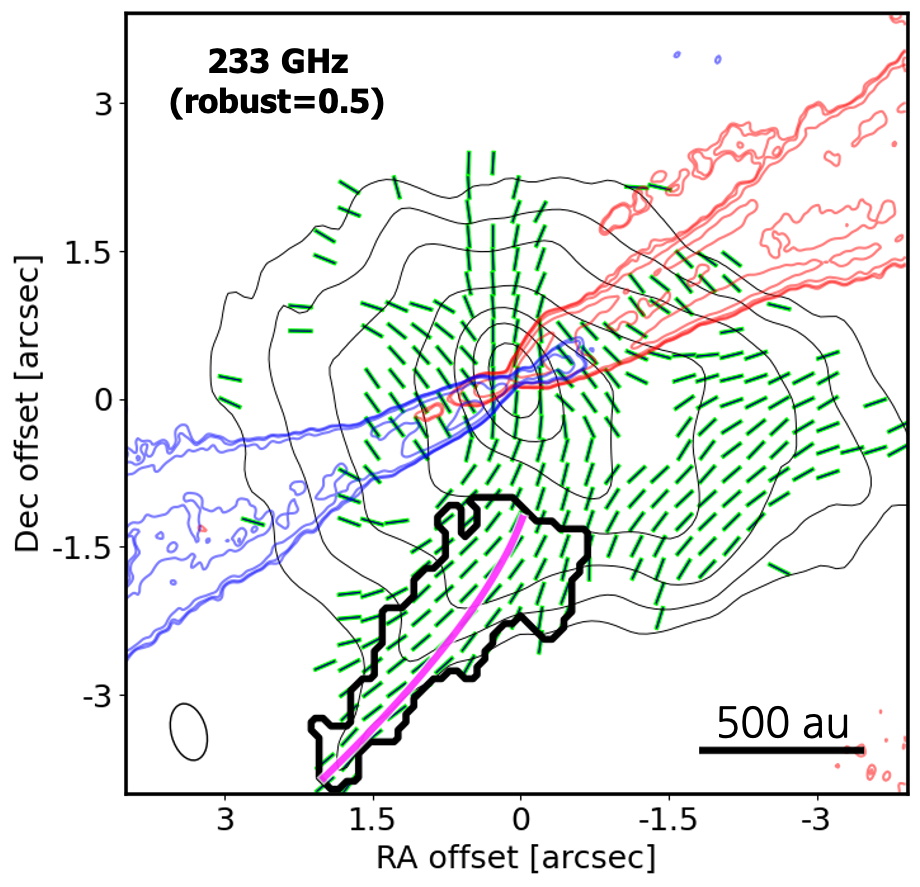}
\caption{{\yw Defined possible southern stream region used to estimate the mass infall rate along the southern filament. The magenta line indicates the ballistic infalling model trajectory within the defined region. All other features are the same as in the right panel of Figure \ref{fig:B_field}.}
\label{fig:stream}}
\end{figure}

{\yc The southern envelope exhibits high column density, and {\yw the magnetic field orientations in this region are consistent with a ballistic infall model.} {\yw This raises the possibility that, if the southern filament indeed traces dense infalling streams along the midplane, such flows could contribute substantially to the overall mass accretion in the HH 211 system.} To quantitatively estimate the mass infall rate along the southern filament, we first define a {\yw possible} southern stream region where the magnetic field orientations are consistent with the modeled ballistic infall trajectories. Figure \ref{fig:stream} shows the defined stream region. We exclude the central part of the envelope ($I > 50\sigma_I$), where the mass is uncertain due to optically thick disk emission being smoothed out by the beam pattern.}

{\yc Using the column density map shown in the right panel of Figure \ref{fig:temp_dens}, the mass of the defined southern stream region is estimated to be $\sim$0.15 $M_\odot$. Alternatively, the mass can be derived directly from the dust continuum emission under the assumption of optically thin thermal emission, using the following relation \citep{Hildebrand_1983}:

\begin{equation}
M = \frac{F_\nu D^2}{\kappa_{\nu} B_{\nu}(T)},
\label{eq:mass2}
\end{equation}
where $F_{\nu}$ is the total flux density, $D$ is the distance to the source, $\kappa_{\nu}$ is the dust opacity, and $B_{\nu}(T)$ is the Planck function at temperature $T$. The integrated flux density within the defined region in Band 6 is 10 mJy. Assuming a dust opacity of $\kappa_{\nu} = 0.01$ cm$^2$ g$^{-1}$ at 233 GHz with a gas-to-dust mass ratio of 100 \citep{Ossenkopf_1994} and a dust temperature in the range $T = 5$–10 K, the inferred total mass is $0.05$–$0.19$ $M_\odot$. It is important to note that the measured dust continuum flux within the defined region may include contributions from background and foreground envelope material in addition to the infalling stream itself. Therefore, the derived mass should be considered an upper limit for the mass of the stream.}

{\yc The length and velocity of the stream within the defined region are estimated based on the CMU model trajectory. The total length of the stream is approximately 6160 au, and the velocity increases along the path from 0.136 km s$^{-1}$ to 0.306 km s$^{-1}$. The average inward velocity along the stream is calculated using $\bar{v} = \int{v \hspace{0.05cm} dl}/L$, where $v$ is the local infall velocity, $dl$ is a differential segment along the trajectory, and $L$ is the total length of the stream. Using this definition, the average velocity is found to be $\bar{v} = 0.188$ km s$^{-1}$. The mass infall rate along the stream is calculated using $\dot{M} = M / (L/\bar{v})$, where $M$ is the mass of the stream. Using the inferred mass, length, and velocity, the estimated mass infall rate along the defined stream region is $(0.32$–$1.22) \times 10^{-6}$ $M_{\odot}$ yr$^{-1}$. Note that the estimated mass infall rate should be considered an upper limit, as the stream mass represents an upper limit.}

{\yc The mass accretion rate onto the the central protostar can be estimated using the following relation:

\begin{equation}
\dot{M}_{\text{acc}} = \frac{L_{\text{acc}} R_{*}}{G M_{*}},
\label{eq:accretion}
\end{equation}
where $L_{\text{acc}}$ is the accretion luminosity, $R_{*}$ is the stellar radius, $M_{*}$ is the stellar mass, and $G$ is the gravitational constant. The bolometric luminosity of the HH 211 system is estimated to be $L_{\text{bol}}\sim$3.6 $L_{\odot}$ \citep{Froebrich_2005}. Assuming an accretion luminosity of $0.5–1$ $L_{\text{bol}}$ and a stellar radius of $1–3$ $R_{\odot}$ \citep{Stahler_1988}, the resulting mass accretion rate is calculated to be $(0.72–4.30) \times 10^{-6}$ $M_{\odot}$ yr$^{-1}$. The mass infall rate along the southern stream, estimated as $(0.32–1.22) \times 10^{-6}$ $M_{\odot}$ yr$^{-1}$, can account for a significant fraction of this accretion rate. {\yw This suggests that if the southern filament indeed represents an infalling stream, it could play a crucial role in supplying mass to the central protostar in the HH 211 system.}}

\subsection{Dynamical Role of Magnetic Fields in the HH 211 System} \label{subsec:role_Bfield}

{\yc On the core scale ($\sim$10,000 au), the mass-to-flux ratio of the HH 211 core is estimated to be 1.2–3.7 \citep{Yen_2023} and 0.45–2.20 \citep{Choi_2024}. Since these estimates are based on plane-of-sky magnetic field measurements and do not account for the line-of-sight component, they should be considered upper limits. Accordingly, the HH 211 core region on this scale is either magnetically subcritical or near critical, where magnetic fields dominates or balances gravity. Figure \ref{fig:jcmt} indeed shows that the core-scale ($\sim$10,000 au) magnetic field orientation is well aligned with the $\sim$4000 au long southern and northwestern filaments. This alignment is unlikely to result from material dragging the field, as the filament widths are much smaller than the core-scale magnetic field structures, and the field strength on this scale is substantial. Instead, the strong magnetic field likely guides the formation of filamentary structures along its direction. Thus, magnetic fields likely play a significant role in regulating the dynamics of the system on the core scale.}

{\yc On the inner envelope scale ($\sim$1000 au), the mass-to-flux ratio of the HH 211 envelope region is estimated to be 9.1–32.3 \citep{Yen_2023}, indicating that gravity dominates over magnetic forces. The right panel of Figure \ref{fig:B_field} shows an hourglass-shaped magnetic field morphology and possible toroidal components along the outflow direction, likely shaped by gas motions dragging the field lines. In the southern envelope, the magnetic field orientation closely follows the expected infall trajectory along the disk midplane. These results suggest that magnetic fields at this scale are not dynamically dominant and instead follow the kinematic motions of the gas.}

{\yc Between the core scale ($\sim$10,000 au) and the inner envelope scale ($\sim$1000 au), magnetic flux must be significantly reduced. This extraction may occur through non-ideal MHD effects such as ambipolar diffusion, in which neutral particles drift relative to the magnetic field lines \citep{Mouschovias_1992}. Another possibility is mass accumulation along magnetic field lines, which increases the local mass without a corresponding increase in magnetic flux \citep{Vazquez_2011}. The alignment of the $\sim$4000 au southern and northwestern filaments with the core-scale magnetic field, together with the inner envelope’s flattened structure perpendicular to the large-scale field, suggests that preferential mass accumulation along field lines could lead to a high mass-to-flux ratio in the inner envelope. This, in turn, would allow infall and rotational motions to become dominant in governing the dynamics of the system. We note that flow along field lines and ambipolar diffusion are not mutually exclusive. They can work together to increase the mass-to-flux ratio at smaller distances.}

\section{Conclusions} \label{sec:conclusion}
{\yc We present ALMA Band 3 and Band 6 polarization observations at $\sim$0.5$^{\prime\prime}$ resolution toward the Class 0 protostellar system HH 211. The Band 3 dust continuum emission reveals three filamentary structures, each approximately 4000 au in length, within the HH 211 envelope. Two of these, the southern and northwestern filaments, are well aligned with the core-scale ($\sim$10,000 au) magnetic fields previously detected with JCMT observations. This alignment suggests that the formation of these filamentary structures may have been influenced by the large-scale magnetic field.}

{\yc The Band 3 polarization observations show that the magnetic field orientations are generally parallel to the southern and western filamentary structures. The magnetic field is not clearly detected in the northwestern filament. On the inner envelope scale ($\sim$1000 au), the Band 6 polarization observations reveal a clear hourglass-shaped magnetic field morphology around the central protostar, and toroidal magnetic fields along the outflow directions. These features suggest that the magnetic fields in the inner envelope are being dragged and shaped by infalling and rotational motions.}

{\yc We estimate the line-of-sight–averaged temperature and column density distributions in the inner envelope of HH 211 using Band 3 and Band 6 continuum emission. The resulting maps reveal asymmetric temperature and density structures. The temperature is higher in the eastern part of the envelope, while the column density is enhanced in the southern and western regions. These high-density regions are aligned with the larger-scale dust filament directions, suggesting that the filaments correspond to high-density regions within the envelope.}

{\yw The southern dense region on a $\sim$1000 au scale exhibits strong polarized intensity and well-aligned magnetic field lines. Since it coincides with the outflow locations, this structure may be outflow cavity walls. Alternatively, this could represent an infalling stream, as the magnetic field orientations are closely aligned with the parabolic trajectories predicted by the ballistic infall model. In this interpretation, we suggest that the magnetic fields in the inner envelope ($\sim$1000 au) may be shaped by infalling motions. If the southern dense region indeed corresponds to an infalling stream, the estimated mass infall rate suggests that it could play a significant role in the overall mass accretion onto the central protostar.}

\vspace{5mm}

We are grateful to the anonymous referee for helpful comments. This work was supported by the National Research Foundation of Korea (NRF) grant funded by the Korea government (MSIT) (RS-2024-00342488 and RS-2024-00416859). L.W.L. acknowledges support from NSF 2307844. Z.Y.L is supported in part by NASA 80NSSC20K0533, NSF AST-2307199, and the Virginia Institute of Theoretical Astronomy (VITA). This paper makes use of the following ALMA data: ADS/JAO.ALMA\#2016.1.00604 and ADS/JAO.ALMA\#2017.1.01310. ALMA is a partnership of ESO (representing its member states), NSF (USA) and NINS (Japan), together with NRC (Canada), MOST and ASIAA (Taiwan), and KASI (Republic of Korea), in cooperation with the Republic of Chile. The Joint ALMA Observatory is operated by ESO, AUI/NRAO and NAOJ. The National Radio Astronomy Observatory is a facility of the National Science Foundation operated under cooperative agreement by Associated Universities, Inc.

\facilities{ALMA, JCMT}
\software{CASA \citep{CASA_2022}, NumPy \citep{Harris_2020}, Matplotlib \citep{Hunter_2007}, Astropy \citep{Astropy_2013,Astropy_2018,Astropy_2022}, SciPy \citep{Virtanen_2020}}

\appendix

\section{Additional polarization properties from the observations}
\label{sec:app}

We present dust continuum toward the HH 211 envelope in Band 6 (233 GHz) with a robust parameter of 2.0. We also present the magnetic field orientations and polarization properties of the HH 211 envelope based on the Band 3 observations with a robust parameter of 0.5 and the Band 6 observations with a robust parameter of 2.0.

\begin{figure*}[ht!]
\epsscale{0.8}
\plotone{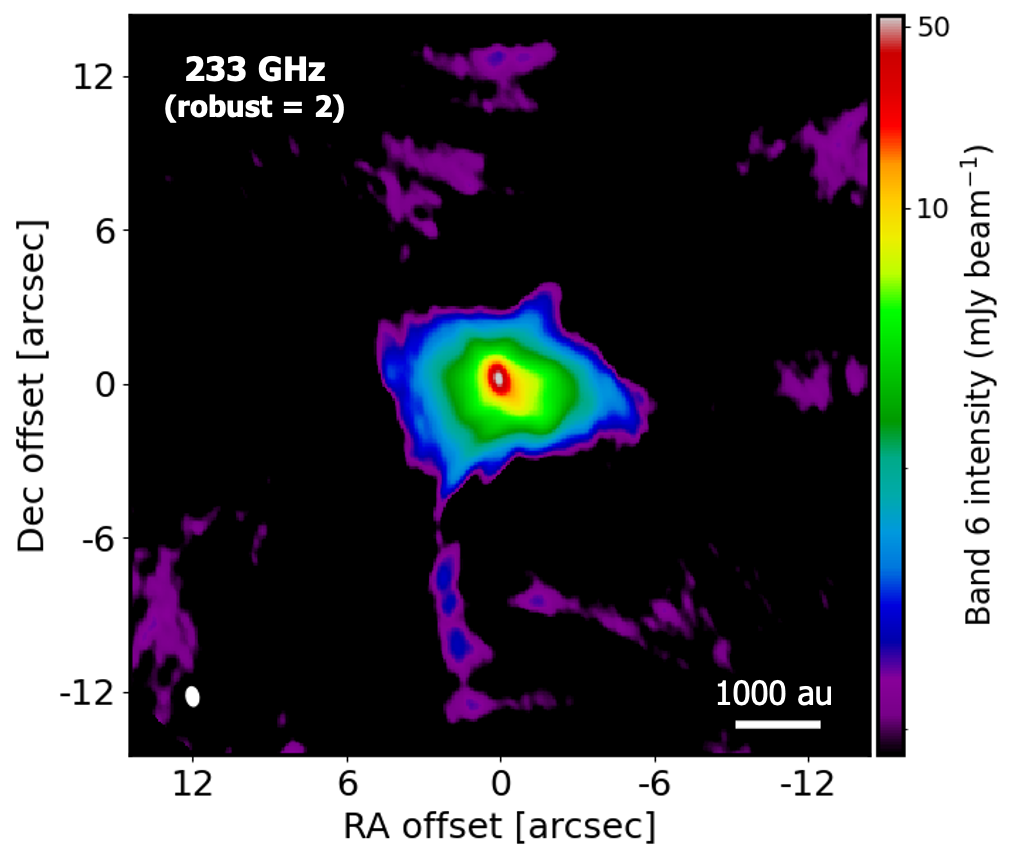}
\caption{Dust continuum around the HH 211 in Band 6 (233 GHz) with a robust parameter of 2.0. The synthesized beam and spatial scale bar are shown in the bottom left and right corners, respectively.
\label{fig:cont_band6}}
\end{figure*}

\begin{figure*}[ht!]
\epsscale{1.15}
\plotone{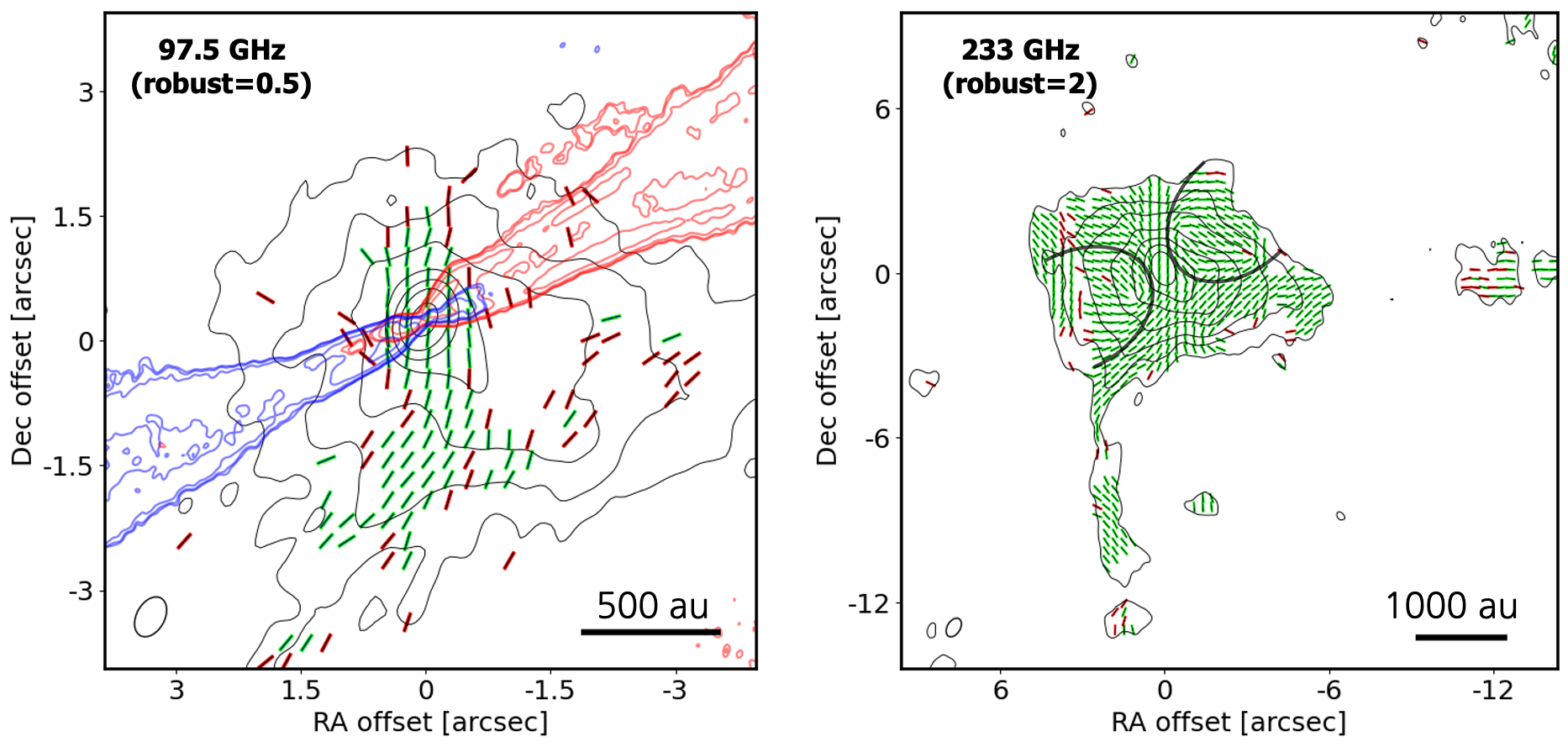}
\caption{Left: Inferred magnetic field orientations around the HH 211 from Band 3 polarized continuum with a robust parameter of 0.5. The line segments represent plane-of-sky magnetic field directions, with green for $PI > 2\sigma_{PI}$, and red for $1.5\sigma_{PI}< PI < 2\sigma_{PI}$. The blue and red contours indicate CO (2-1) integrated intensities in velocity ranges from 0 to 8 and from 10 to 18 km s$^{-1}$, respectively. The blue and red contour levels are 0.03, 0.035, 0.05, 0.1, and 0.2 K km s$^{-1}$. The black lines are the Band 3 dust continuum with a robust parameter of 0.5, showing 5, 10, 20, 50, 100, 200, and 500 times the rms noise level. Right: Inferred magnetic field orientations around the HH 211 from Band 6 polarized continuum with a robust parameter of 2.0. Green segments represent plane-of-sky magnetic field directions with green for $PI > 2\sigma_{PI}$, and red for $1.5\sigma_{PI}< PI < 2\sigma_{PI}$. The black lines are the Band 6 dust continuum with a robust parameter of 2.0, showing 5, 10, 20, 50, 100, 200, and 500 times the rms noise level. The gray U-shaped curves indicate the pinched magnetic field patterns around the protostar. The synthesized beam size and spatial scale bar are shown in the bottom left and right of both panels, respectively.
\label{fig:B_field_appendix}}
\end{figure*}

\begin{figure*}[ht!]
\epsscale{1.15}
\plotone{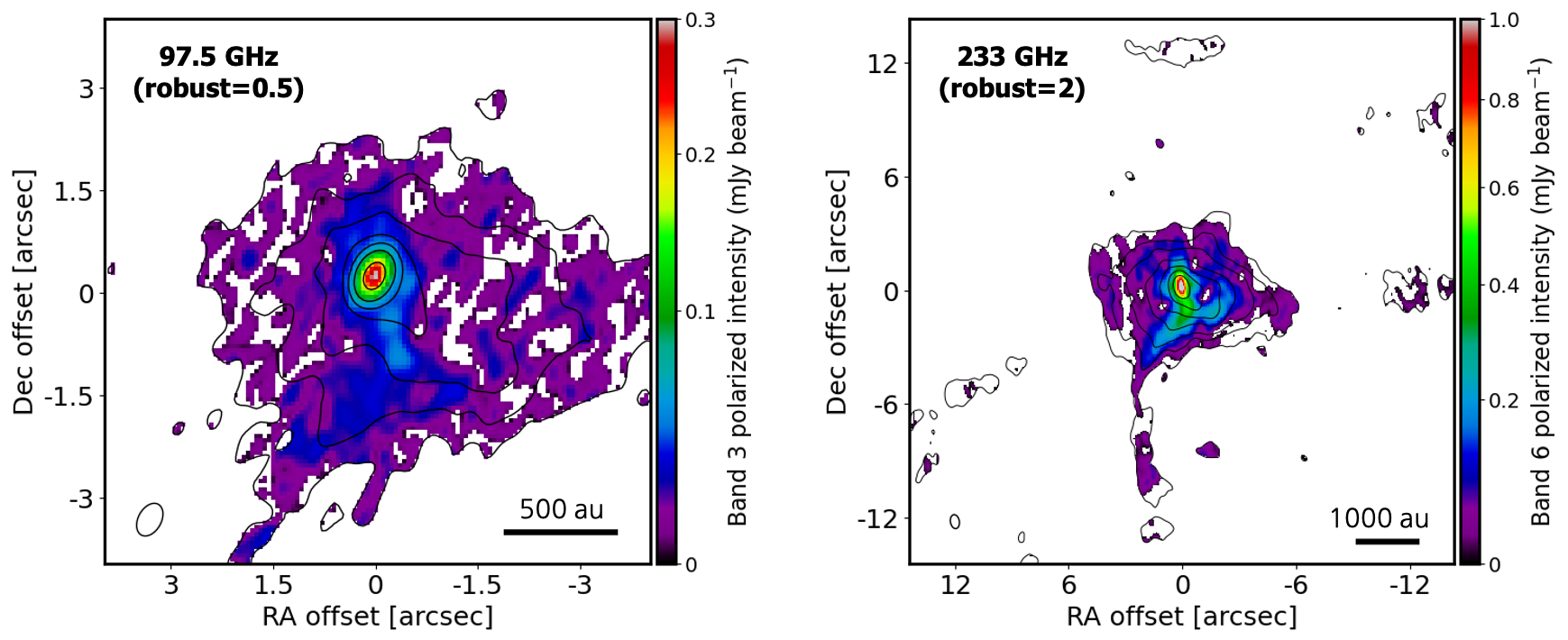}
\caption{Polarized intensity around the HH 211 envelope in Band 3 with a robust parameter of 0.5 (Left) and in Band 6 with a robust parameter of 2.0 (Right). The black lines represent Stokes $I$ intensity levels with 5 (Left) / 3 (Right), 10, 20, 50, 100, 200, and 500 times the rms noise level in each observation. The synthesized beam size and spatial scale bar are shown in the bottom left and right of both panels, respectively.
\label{fig:pol_I_appendix}}
\end{figure*}

\begin{figure*}[ht!]
\epsscale{1.15}
\plotone{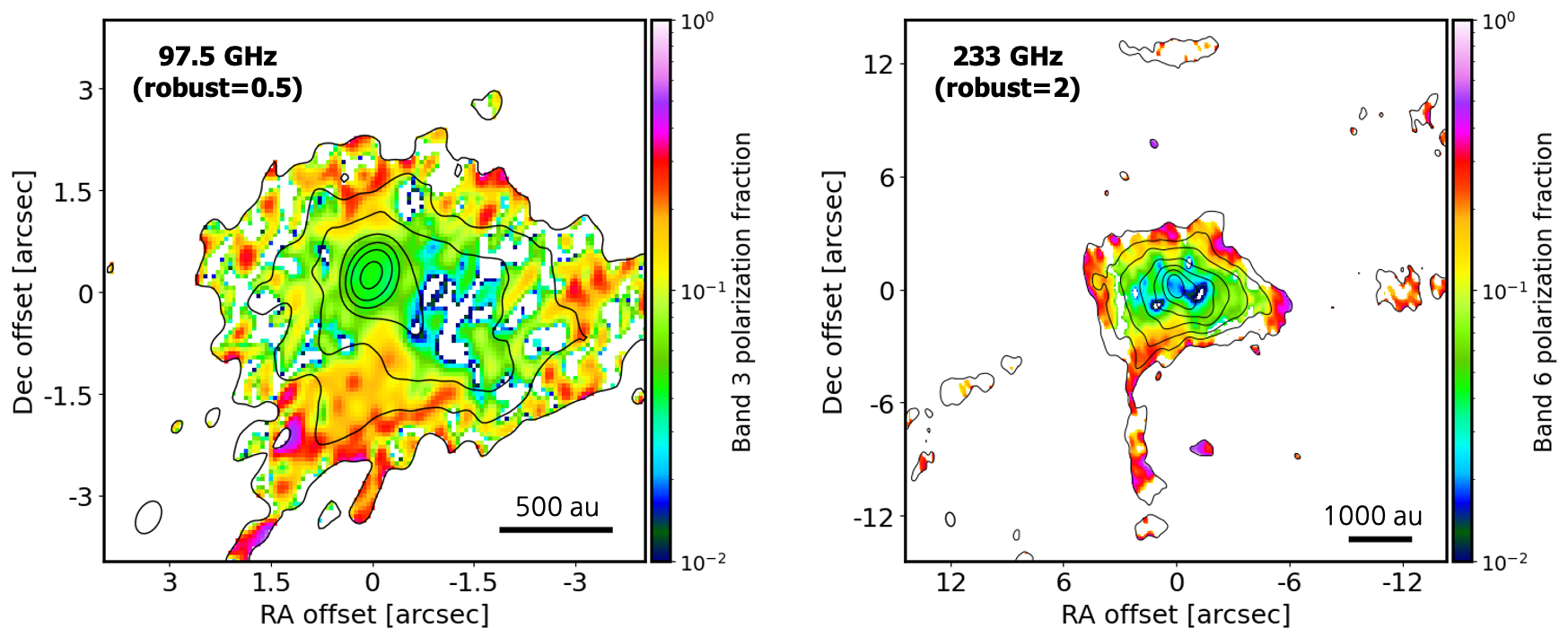}
\caption{Polarization fraction around the HH 211 envelope in Band 3 with a robust parameter of 0.5 (Left) and in Band 6 with a robust parameter of 2.0 (Right). The black lines represent Stokes $I$ intensity levels with 5 (Left) / 3 (Right), 10, 20, 50, 100, 200, and 500 times the rms noise level in each observation. The synthesized beam size and spatial scale bar are shown in the bottom left and right of both panels, respectively.
\label{fig:pol_frac_appendix}}
\end{figure*}

\section{Modeling the HH 211 envelope} \label{sec:modeling}
We make models and carry out synthetic observations to understand the observed spectral index structure, the estimated line-of-sight temperature, and the column density distributions discussed in Section \ref{sec:discussions}. The goal of our modeling is not to derive the best model parameters for the HH 211 envelope. Rather, it is to assess the reliability of our analysis in Section \ref{sec:discussions}.

We construct models with an embedded disk and a flattened protostellar envelope. The radius of the disk in our model is 20 au and the height is 10 au with a position angle of $54^{\circ}$ and an inclination of $81^{\circ}$ \citep{Lee_2018b,Lee_2019}. We assume uniform density and temperature of the disk since it is not resolved in our observations. We take $\rho_{\text{disk}}=3.0 \times 10^{-14}$ g cm$^{-3}$ , and $T_{\text{disk}} =100$ K \citep{Yen_2023}. The density and temperature distributions of the model envelope are as follows:

\begin{equation}
\rho_{\mathrm{env}} = \rho_0 \bigg(\frac{r^{\prime}}{r_0}\bigg)^{p},
\label{eq:dens}
\end{equation}

\begin{equation}
T_{\mathrm{env}} = T_0 \bigg(\frac{r^{\prime}}{r_0}\bigg)^{q},
\label{eq:temp}
\end{equation}

\begin{equation}
r^{\prime} = \sqrt{x^2 + y^2 + (z/f)^2}.
\label{eq:radius}
\end{equation}

\begin{deluxetable*}{lll}
\tablenum{1}
\tablecaption{The Model Parameters \label{tab:parameters}}
\tablewidth{0pt}
\tabletypesize{\footnotesize}
\tablehead{\colhead{Parameter} & \colhead{Description} & \colhead{Value}}
\startdata
\multicolumn{3}{c}{\textbf{Disk Parameters}} \\
$\rho_{\mathrm{disk}}$ & Dust mass density in the disk & $3.0\times 10^{-14}$ g cm$^{-3}$ \\
$T_{\mathrm{disk}}$ & Temperature in the disk & 100 K \\
$r_{\mathrm{disk}}$ & Radius of the disk & 20 au\\
$h_{\mathrm{disk}}$ & Height of the disk & 10 au\\
\hline
\multicolumn{3}{c}{\textbf{Envelope Parameters}} \\
$\rho_{0}$ & Dust mass density at a radius of 100 au & $3.0\times 10^{-17}$ g cm$^{-3}$\\
$T_{0}$ & Temperature at a radius of 100 au & 20 K\\
$p$ & Power-law index of the density profile & -1.5\\
$q$ & Power-law index of the temperature profile & -0.65\\
$f$ & Envelope flattening parameter & 0.7\\
\hline
\multicolumn{3}{c}{\textbf{Filament Parameters}} \\
$\Delta \rho_{\mathrm{fil}}$ & Filament mass density & $6.0\times 10^{-18}$ g cm$^{-3}$\\
$\Delta T_{\mathrm{fil}}$ & Temperature enhancement & 6 K\\
$r_{\mathrm{fil}}$ & Filament width & 200 au\\
\enddata
\label{tab:parameters}
\end{deluxetable*}

Both the density and temperature have power-law profiles in our model. We insert the parameter $f = 0.7$ in the $z$ direction to make a flattened envelope. Following the results of \cite{Yen_2023}, we define $r_0 = 100$ au and set $\rho_0$ = $3.0\times 10^{-17}$ g cm$^{-3}$, $p=-1.5$, $T_{0}$ = 20 K, $q=-0.65$. The power-law slope of the temperature $q$ in \cite{Yen_2023} is -0.4, but we set $q$ to be -0.65 to match the observational features. The model parameters are summarized in Table \ref{tab:parameters}.

\begin{figure*}[ht!]
\epsscale{1.14}
\plotone{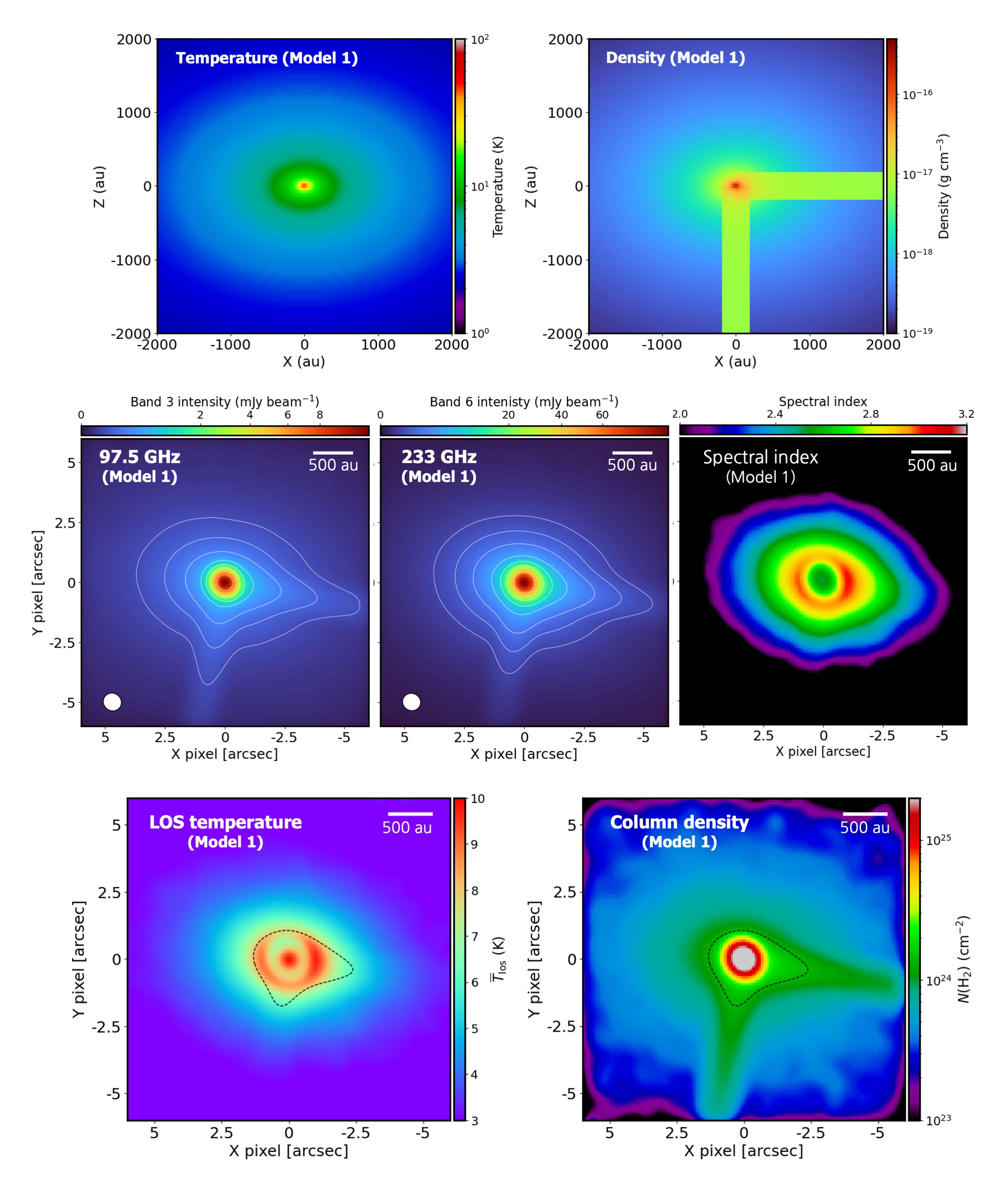}
\caption{The density-enhanced model. Top:  Temperature and dust mass density in the $x-z$ plane ($y=0$). Middle: Synthetic images in Band 3 (97.5 GHz) and Band 6 (233 GHz), and the spectral index from the two synthetic images. Contour levels are 5, 10, 20, 50, 100, 200, and 500 times of the rms noise level in the observations. Bottom: The line-of-sight averaged temperature and the column density calculated from the two synthetic images. The black dotted line indicates the optical depth of 0.1 at 233 GHz. Note that the estimated values are uncertain in the central region.
\label{fig:model1}}
\end{figure*}

\begin{figure*}[ht!]
\epsscale{1.14}
\plotone{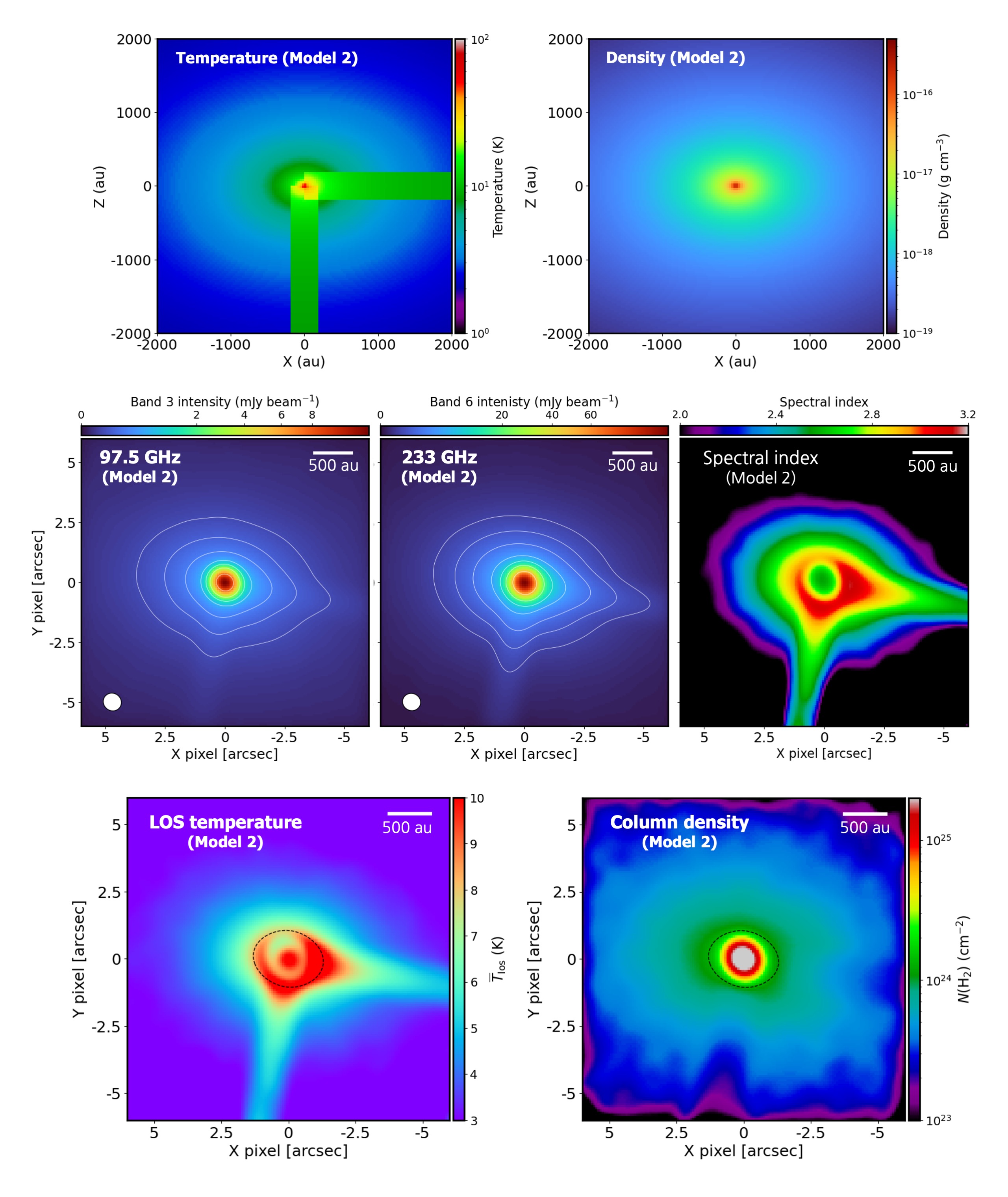}
\caption{The temperature-enhanced model. Individual panels are same to Figure \ref{fig:model1}.
\label{fig:model2}}
\end{figure*}

We construct two models with filament structures. The first model (Model 1) features a protostellar envelope with dense filaments. In this model, the density is high along the filaments, but the temperature remains consistent with that of the envelope. In contrast, the second model (Model 2) has high temperature along the filaments, corresponding to enhanced temperatures along the outflow cavity wall. Comparing these models to the observations allows us to determine whether the observed asymmetric dust continuum and spectral index are due to dense filaments or high temperatures. We consider two cylindrical filaments: one to the west and one to the south. The filament regions are defined as follows:

\begin{equation}
\mathrm{Filament} \hspace{0.05cm} 1: x>0, \hspace{0.15cm} y^2+z^2<r_{\mathrm{fil}}^2,
\label{eq:fila1}
\end{equation}

\begin{equation}
\mathrm{Filament} \hspace{0.05cm} 2: z<0, \hspace{0.15cm} x^2+y^2<r_{\mathrm{fil}}^2,
\label{eq:fila2}
\end{equation}
where $r_{\mathrm{fil}}$ represents the filament width, measured as a radius. We set $r_{\mathrm{fil}}$ to 200 au. 
For the density-enhanced model (Model 1), we add an additional density $\Delta \rho_{\mathrm{fil}}$ to the filament regions. The dust mass density of the filament is set to $\Delta \rho_{\mathrm{fil}}$ = $6.0\times 10^{-18}$ g cm$^{-3}$. In the temperature-enhanced model (Model 2), we add an additional temperature $\Delta T_{\mathrm{fil}}$ to the filament regions. The additional temperature of the filament regions is set to $\Delta T_{\mathrm{fil}}$ = 6 K. The top panels of Figures \ref{fig:model1} and \ref{fig:model2} show the model temperature and density distributions in the $x-z$ plane ($y=0$).

We run RADMC-3D to generate synthetic images in Band 3 (97.5 GHz) and Band 6 (233 GHz) frequencies \citep{Dullemond_2012}. We use Optool to make dust opacity tables \citep{Dominik_2021}. The maximum sizes of dust particles are assumed to be 200 $\mu$m in the disk and 10 $\mu$m in the envelope and the filaments. Dust particles have thin icy mantles with MRN distributions in our models. The position angle of the envelope is assumed to be $10^{\circ}$. After making synthetic images in Band 3 and Band 6 frequencies, we calculate the spectral index, the line-of-sight averaged temperature, and the column density using the same methods described in Sections \ref{sec:results} and \ref{sec:discussions}.

The middle and bottom panels of Figure \ref{fig:model1} show the synthetic images, the spectral index, the line-of-sight averaged temperature and the column density from the density-enhanced model (Model 1). The western and southern dust filaments are seen in the synthetic images. However, the filamentary structures are not apparent in the spectral index map, which resembles the observed spectral index map. This is because, in the optically thin limit, the density does not affect the spectral index. The general structure of the synthetic spectral index map is similar to the observed spectral index. It reproduces the low spectral index in the center and the decreasing trend outward. The inward decreasing pattern in the central region is attributed to the optically thick disk emission. The black dotted lines in the bottom panels indicate an optical depth of 0.1 at 233 GHz, where spectral index in dust opacity drops abruptly (see the right panel of Figure \ref{fig:alpha}). The outward decreasing trend is due to the lower temperature. The estimated line-of-sight averaged temperature and the column density clearly identify the high-density filaments in the south and west. As with the observations, the estimated values are uncertain in the central region.

The middle and bottom panels of Figure \ref{fig:model2} show the synthetic images, the spectral index, the line-of-sight averaged temperature and the column density from the temperature-enhanced model (Model 2). The western and southern temperature-boosted regions are seen in the synthetic images. In contrast to the density-enhanced model, the filamentary structures are stronger in the Band 6 image than the Band 3 image since the change in intensity with temperature is more sensitive at higher frequencies. The synthetic spectral index map shows high spectral index along the filaments due to high temperatures in the non-Rayleigh-Jeans realm. The estimated line-of-sight averaged temperature and the column density clearly identify the high-temperature regions, though there is some distortion in the column density in the south and west. Overall, the general observed features are more similar to those predicted by the density-enhanced model than by the temperature-enhanced model. Therefore, we suggest that the HH 211 envelope contains dense filaments in the south and the west.

\section{Inferring the CMU Model Trajectory} \label{sec:cmu}

{\yc The procedure for inferring the CMU model trajectory is as follows. First, we compute the position ($r, \theta, \phi$) and velocity components ($v_r, v_\theta, v_\phi$) in spherical coordinates using Equations (5)–(10) from \citet{Ulrich_1976}. These quantities are then converted into Cartesian coordinates, where the disk midplane lies in the $x-y$ plane and the angular momentum vector is aligned with the $+z$-axis. The initial polar angle, $\theta_0$, ranges from $0^\circ$ to $180^\circ$, with $\theta_0 = 0^\circ$ corresponding to a particle initially located along the angular momentum axis, and $\theta_0 = 90^\circ$ corresponding to a particle starting in the disk midplane. The initial azimuthal angle, $\phi_0$, ranges from $0^\circ$ to $360^\circ$.

To transform from the disk frame to the observer frame, where the new $x-y$ plane corresponds to the plane of the sky and the new $z$-axis points along the line of sight, we apply a sequence of rotations. First, we rotate about the $x$-axis by the disk inclination angle $i$. We define $i = 0^\circ$ as a face-on disk with clockwise rotation, which naturally aligns the disk $x-y$ plane with the plane of the sky and the $+z$-axis with the line of sight. The inclination increases as the $x$-axis rotates from the $y$-axis toward the $z$-axis, with $i = 90^\circ$ corresponding to an edge-on disk in which the eastern side (positive $x$) is redshifted. This definition may differ from conventional definitions. The inclination $i$ ranges from $0^\circ$ to $360^\circ$.

After applying the inclination rotation about the $x$-axis, we rotate the system around the $z$-axis by the position angle (PA) of the disk. The PA ranges from $-90^\circ$ to $90^\circ$, where a disk major axis aligned with the $x$-axis corresponds to PA = $0^\circ$, and a major axis oriented in the northeast–southwest direction corresponds to a positive PA. After these two rotations, the position and velocity vectors are fully transformed into the observer’s frame. The resulting $x$ and $y$ components represent the trajectory in the plane-of-sky, while the $v_z$ component corresponds to the line-of-sight velocity.}

\clearpage

\bibliographystyle{aasjournalv7}
\bibliography{mybib}

\end{document}